\documentclass[twocolumn]{emulateapj}  
\usepackage{times}

\usepackage{graphicx,bm,amssymb}
\usepackage{ulem}
\usepackage{epsfig}
\usepackage{amssymb}
\usepackage{natbib}
\usepackage{pstricks}
\usepackage{psfrag}
\usepackage[colorlinks=true,linkcolor=blue,citecolor=blue]{hyperref}
\def\aj{AJ}%% Astronomical Journal
\def\araa{ARA\&A}%% Annual Review of Astron and Astrophys
\def\apj{ApJ}%% Astrophysical Journal
\def\apjl{ApJ}%% Astrophysical Journal, Letters
\def\apjs{ApJS}%% Astrophysical Journal, Supplement
\def\apss{Ap\&SS}%% Astrophysics and Space Science
\def\aap{A\&A}%% Astronomy and Astrophysics
\def\aapr{A\&A~Rev.}%% Astronomy and Astrophysics Reviews
\def\aaps{A\&AS}%% Astronomy and Astrophysics, Supplement
\def\caa{Chinese Astron. Astrophys.}%% Chinese Astronomy and Astrophysics
\def\mnras{MNRAS}%% Monthly Notices of the RAS
\def\nar{New A Rev.}%% New Astronomy Review
\def\pasp{PASP}%% Publications of the ASP
\def\pasj{PASJ}%% Publications of the ASJ
\def\sovast{Soviet~Ast.}%% Soviet Astronomy
\def\nat{Nature}%% Nature
\newcommand{\be}{\begin{equation}}
\newcommand{\ee}{\end{equation}}
\newcommand{\bary}{\begin{eqnarray}}
\newcommand{\eary}{\end{eqnarray}}

\shorttitle{Long-term optical polarization of Mrk\,421}
\shortauthors{Fraija N. et al.}
\begin{document}
\title{Long-term optical polarization variability and multiwavelength analysis of Blazar Mrk\,421}
\author{N. Fraija\altaffilmark{1}$^\dagger$, E. Ben{\'i}tez \altaffilmark{1},  D. Hiriart \altaffilmark{2,3}, M. Sorcia\altaffilmark{1}, J. M. L\'opez\altaffilmark{4},  R. M\'ujica\altaffilmark{5}, J. I. Cabrera\altaffilmark{6}, J. A. de Diego\altaffilmark{1}, 
M. Rojas-Luis\altaffilmark{1},  F. A. Salazar-V\'azquez \altaffilmark{1}  and  A. Galv\'an-G\'amez \altaffilmark{1}} 
\altaffiltext{1}{Instituto de Astronom\'ia, Universidad Nacional Aut\'onoma de M\'exico, Apdo. Postal 70-264, 04510 Cd. de M\'exico, Mexico.\\ $\dagger$ \email{nifraija@astro.unam.mx}}
\altaffiltext{2}{Instituto de Astronom\'ia, Universidad Nacional Aut\'onoma de M\'exico, Ensenada, Baja California, Mexico}
\altaffiltext{3}{Visiting professor K\"{o}nigstuhl Landessternwarte (LSW) Heidelberg University, Germany}
\altaffiltext{4}{Facultad de Ciencias, Universidad Aut\'onoma de Baja California, Campus El Sauzal, Ensenada B.C., Mexico.}
\altaffiltext{5}{Instituto Nacional de Astrof\'isica, \'Optica y Electr\'onica, Apdo. Postal 51 y 216, 72000 Tonantzintla, Puebla, Mexico}
\altaffiltext{6}{Facultad de Ciencias, Universidad Nacional Aut\'onoma de M\'exico, 04510 Cd de M\'exico, Mexico}
  \date{\today}

\begin{abstract}

The results of 8-year R-band photopolarimetric data of blazar Mrk\,421 collected from February 2008 to May 2016  are presented, along with
extensive multiwavelength observations covering from radio to TeV $\gamma$-rays around the flares observed in May 2008, March 2010, and April 2013. 
The most important results are found in 2013 when the source displayed in the R-band a very high brightness state of $11.29\pm 0.03$ mag ($93.60\pm  1.53$ mJy) on April 10th and a polarization degree of ($11.00\pm0.44$)\% on May 13th. The analysis of the optical data shows that the polarization variability  is due to the superposition of two polarized components that might be produced in two distinct emitting regions.  An intranight photopolarimetric variability study carried out over 7 nights after the 2013 April maximum found flux and polarization variations on the nights of April 14, 15, 16 and 19.  In addition, the flux shows a minimum variability timescale of  $\Delta\,t=$2.34$\pm$0.12 hours, and that the polarization degree presented variations  of $\sim$ (1 - 2) \% in a timescale of $\Delta\,t\sim$ minutes.  Also, a detailed analysis of the intranight data shows a coherence length of the large-scale magnetic field of $l_B\simeq$ 0.3 pc which is the same order of magnitude to the distance traveled by the relativistic shocks.  This result suggests that there is a connection between the intranight polarimetric variations  and spatial changes of the magnetic field. Analysis of the complete R-band data along with the historical optical light curve found for this object shows that Mrk\,421 varies with a period of 16.26 $\pm$ 1.78 years.

\end{abstract}


\section{Introduction}

Blazars, a special subclass of active galactic nuclei (AGN) are characterized by having a relativistic jet closely aligned with the observer's line of sight (estimated viewing angles $\theta\lesssim 10^{\circ}$).   Blazars exhibit high variability in all frequencies of the electromagnetic spectrum. Several variability studies in blazars  have been  carried out with the aim of finding short and long timescale variabilities \citep{1997ARA&A..35..445U,2014A&ARv..22...73F}.  For instance, optical bands have been well studied to search for variations down to minutes \citep[the so-called intraday variability; see, e.g., ][]{1995ARA&A..33..163W} and periodic or quasi-periodic variations on timescales of years \citep[see, e.g.,][]{1988ApJ...325..628S}.   Strong and variable polarization degree higher than 3\% has been observed in blazars in the radio and optical bands \citep{1980ARA&A..18..321A,1990ApJ...354..124I}.  Polarization variability studies can provide valuable information about the structure and the strength of the magnetic field associated with the physical processes in the emitting region \citep[see, e.g.][]{2014ApJ...794...54S, 2013ApJS..206...11S}.    

Analysis of the equivalent width of the emission lines revealed that blazars can be divided in BL Lac objects and flat spectrum radio quasars \citep[FSRQs;][]{1996MNRAS.281..425M,2012agn..book.....B}.  An additional analysis of the spectral features carried out by \cite{2009ApJ...700..597A} showed a fairly different separation between FSRQ and BL Lac objects, with FSRQs having considerably softer spectra.  They found that the spectral power-law index $\sim$ 2.2 corresponds to the boundary between BL Lac objects and FSRQs.  Later, \cite{2009MNRAS.396L.105G} proposed that this division can be explained in terms of the different accretion regimes and the distinct radiation cooling mechanisms of electrons in these objects.  It is worth noting that according to the location of the synchrotron peak in the spectral energy distribution (SED),  BL Lac objects are classified into low-energy peaked BL Lac (LBL), intermediate-energy peaked BL Lac (IBL) and high-energy peaked BL Lac \citep[HBL;][]{1995ApJ...444..567P}.

At a distance of 134.1 Mpc, the BL Lac object Mrk\,421 \cite[z=0.03;][]{2005ApJ...635..173S} is one of the closest and most comprehensively studied sources of the HBL class. Mrk\,421 was the  first extragalactic object observed in the very-high-energy (VHE) $\gamma$-ray band  \citep{1992Natur.358..477P}.  By analyzing the historical light curve comprising data for almost 100 years in the optical B band, \cite{1997A&AS..123..569L} found two kinds of variability behaviors in Mrk\,421. The first one, consisted of non-periodical rapid variations on timescales from hours to days.  The second one consisted of periodic variations with possible periods of  $23.1\pm1.1$ and $15.3\pm0.7$ years, although the former period was not very significant. 

Considered as an excellent candidate to study the physical processes within blazar jets, Mrk\,421 has been a frequent target of multiwavelength campaigns in order to study correlations among distinct energy bands.    Correlations between TeV $\gamma$-ray and X-ray bands have been found several times during high activity states or flares \citep[][]{1995ApJ...449L..99M,2004NewAR..48..419F, 2007ApJ...663..125A} and also during quiescent states \citep{2015A&A...576A.126A, 2016ApJ...819..156B}. In only a few cases  TeV $\gamma$-rays without an X-ray emission counterpart have been reported \citep[the so-called orphan flares; see, e.g.][]{2005ApJ...630..130B,2009ApJ...703..169A}.   Correlations found between optical bands and TeV $\gamma$-rays/X-rays are still controversial.  For instance, optical and TeV $\gamma$-ray/X-ray correlations and anti-correlations have been found with different time lags in some studies \citep{2009ApJ...695..596H, 2015A&A...576A.126A}  and in other ones no correlations have been reported  \citep{1995ApJ...449L..99M,2015A&A...576A.126A,2007ApJ...663..125A}.  Similarly, the radio bands are found not to be correlated with  $\gamma$-rays  \citep{2011ApJ...738...25A}.

The broadband SED of Mrk\,421 presents a double-humped shape; the lower energy hump has a peak located at a few keV and the second hump at hundreds of GeV.  \cite{2011ApJ...736..131A} found that both leptonic and hadronic models are able to fit  reasonably well the SED of this object, implying comparable jet powers but with very different characteristics for the emitting region.   In the Leptonic scenario, a one-zone synchrotron self-Compton (SSC) model with three accelerated electron power-law functions (through diffusive relativistic shocks with a randomly oriented magnetic field)  has been used \citep{2011ApJ...736..131A}.   In the hadronic scenario \citep{2011ApJ...736..131A, 2015APh....70...54F}, the first peak is explained by electron synchrotron radiation. For  the second peak that extends from low-energy $\gamma$-rays to VHE $\gamma$-rays, the synchrotron proton blazar (SPB) model is used \citep{2001APh....15..121M,2003APh....18..593M}. It is worth noting that the strength of magnetic field required to fit the SED with a hadronic model was 50 G.  On the other hand,  using the shortest variability timescale of 1 hour, a magnetic field density of  $8.2 \times 10^{-2}$ G was obtained when a leptonic model was considered.

The polarization observations of Mrk\,421 have been widely studied with different optical bands.  \cite{1983Afz....19..199H} studied  the polarization and photometric behavior  in the B and V bands during 1974 - 1982. They reported a variability timescale  from days to years and polarization degrees less than 6\% with a preferential position angle of 173$^\circ$. Successively, many photopolarimetric observations were carried out from the ultraviolet to near infrared regions \citep{1990A&AS...83..183M, 1991A&AS...90..161T, 1992A&AS...94...37T, 1993Ap&SS.206..191T}, reporting a moderate polarization degree.   From October 1994 to June 1997,  polarimetric observations were performed in radio and optical bands by \cite{1998A&A...339...41T}. In particular, a large optical outburst was detected  in the winter 1996-97,  followed by a radio outburst with a delay of 30-60 days. At the same time, Mrk\,421 increased its polarization level reaching a value of $\sim$ 12\% in the V band.    \cite{2011PASJ...63..639I} reported correlations between the flux, color (K, J and Ks bands) and polarization variations on timescales from days to months. Recently, using the data collected with KVA and RINGO2 from 2008 to 2011 \cite{2016MNRAS.462.4267J} reported a strong increase in the polarization degree and a 360$^\circ$ rotation of the position angle before the unprecedentedly large $\gamma$-ray flare occurred after June 2012.

In this work, we report  the results of a long-term R-band photopolarimetric observations of blazar Mrk\,421 carried out from  February 2008 to May 2016.    The paper is arranged as follows: in Section 2, the R-band optical polarimetric observations and data reduction are presented. In Section 3, the polarimetric variability analysis is shown. Section 4 describes the multiwavelength data used in this work. Section 5 shows the results of the analysis done to the multiwavelength data. Section 6 presents general results and Section 7 the conclusions.

\section{Optical polarimetric observations and data reduction}

The optical R-band observations were performed with the 0.84 m f/15 Ritchey-Chr\'etien telescope at the Observatorio Astron\'omico Nacional of San Pedro M\'artir (OAN-SPM) in Baja California, Mexico\footnote{A detailed description of our photopolarimetric monitoring program on  blazars can be found in http://www.astrossp.unam.mx/blazars.}  and with the instrument POLIMA, a single beam polarimeter\footnote{The technical information about the POLIMA instrument can be found in http://haro.astrossp.unam.mx/blazars/instrument/instrument.html }. The exposure time was 60 sec per frame of Mrk\,421. For these observations three different CCD cameras were used with a pixel size of 13.5 to 24 $\mu$m, a 2$\times$2 binning and plate scale of 0.22-0.39\,\arcsec$^{-1}$.  Photometry was performed with an aperture radius of $\sim$\,6\arcsec after regular subtraction of bias, dark current, and flat--field correction for each angular position of the polarizer. Images had an average FWHM of $\sim$3\arcsec. Photometric calibrations were done using nearby standard stars from \citet{1996A&AS..116..403F}. Polarimetric calibrations were done using the polarized and unpolarized standard stars from \citet{1992AJ....104.1563S}. 
 A correction for polarization bias was done using the estimator of the polarization amplitude from normally distributed Stokes parameters \citep{2014MNRAS.439.4048P}. The R-band magnitudes were converted to apparent flux using $F_{R}=K_{0}\times\,10^{-0.4m_{R}}$ with K$_{0}=3.08\times10^{6}$ mJy for the effective wavelength of $\lambda$=6400\AA. The host galaxy contribution was removed from  the total and polarized fluxes after fitting its surface brightness profile given by \citet{1968adga.book.....S,1993MNRAS.265.1013C}

\begin{equation}
I(r)=I(r_e){\rm dex} \left\{ -b_{n}\left[\left(\frac{r}{r_e}\right)^{-n} - 1\right]\right\}\,,
\end{equation}

where $b_{n}=0.868n^{-1}-0.142$ and $I(r_{e})=f_R/[K_{n}r_e^2(1-\epsilon)]$ with
$f_R$ the flux corresponding to a total magnitude and  ${\rm log}\, K_{n}=0.03\, [{\rm log}(1/n)]^2+0.441\,  {\rm log} (1/n)+1.079$. Using a a S\'ersic index of n$=$4, an ellipticity of $\epsilon=0.21$ and a diameter of aperture of 7.5$\arcsec$, we found a value of 14.48$\pm$0.18 for the magnitude of the host galaxy in the R-band with an effective wavelength of $\lambda=$6400\,\AA\,  \citep[see, ][]{1999PASP..111.1223N,2013ApJS..206...11S}.  It is worth noting that the correction of the flux from the host galaxy depends strongly on the aperture radius used in the photometry but the FHWM from seeing has only a minor effect \citep{2007A&A...475..199N}.

The position angle was corrected by an ambiguity of 180$\degr$ in such a way that the variations observed between the position angle of temporal consecutive data should be less than 90$\degr$.  This difference can be written as

\begin{equation}
 |\Delta\theta_{m}|=|\theta_{m+1}-\theta_{m}| - \sqrt{\sigma(\theta_{m+1})^2+\sigma(\theta_{m})^2},
\end{equation}

where $\theta_{m}$  and $\theta_{m+1}$ are the  m-th  and  $m+1$ position angles, respectively, and $\sigma(\theta_{m})$ and $\sigma(\theta_{m+1})$ their errors.  If $\Delta\theta_{m}\,<\,-90^{\circ}$,  $180^{\circ}$ are added to $\theta_{m+1}$. If $\Delta\theta_{m}\,>\,90^{\circ}$, $-180^{\circ}$ are added to $\theta_{m+1}$. If $|\Delta\theta_{m}|\leq90^{\circ}$, no correction is needed.

The R-band photopolarimetric observations of the BL Lac Mrk\,421 were carried out since  February 2008 up to May 2016  ( data are available online, see Table \ref{all_data}). The total collected data during 8 years were 385  points, obtained in 30 observing runs.

\begin{figure*}
\centering
\includegraphics[width=0.95\textwidth]{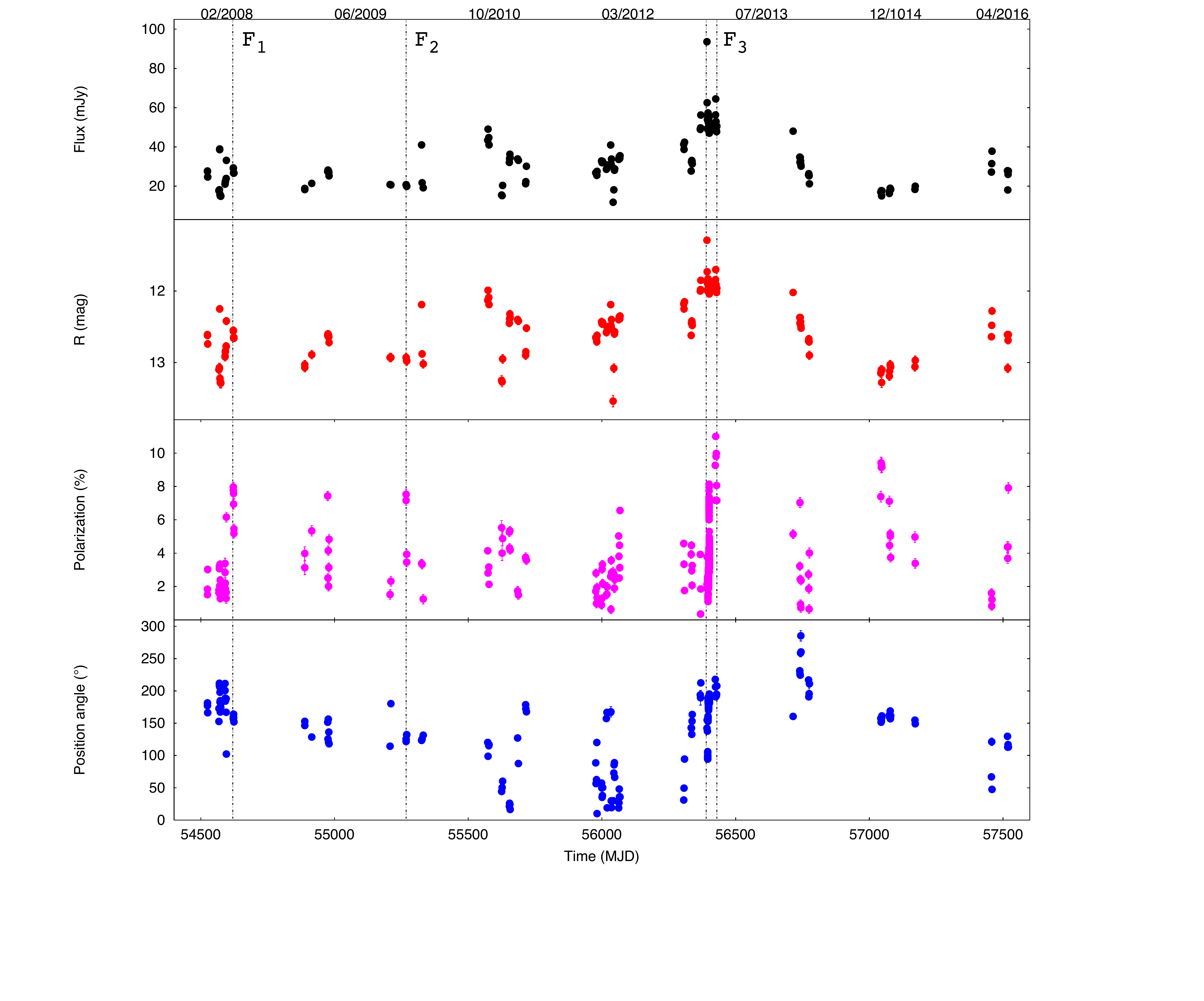}
\caption{OAN-SPM photometric light curve of Mrk\,421 from 2008 February 28 (MJD 54524) to 2016 May 11 (MJD 57519) including polarimetric data are shown. From top to bottom:  The optical flux, the R-band mag, the polarization degree and the position angle variations are presented. Dashed vertical lines mark the observed flares.  F1 represents the flare observed in May 2008,  F2 represents the flare in March 2010 and F3 represents the flare in April 2013.}
\label{optical_all}
\end{figure*}

Figure \ref{optical_all} shows the R-band photopolarimetric lightcurves of  Mrk\,421 obtained since February 2008 up to May 2016. The upper two panels show flux and R-band magnitude variation and the lower two panels show the polarization degree and the  electric vector position angle (EVPA) variations. Hereafter, instead of EVPA, this parameter will be denoted as position angle. Table \ref{gen_data} shows the maximum and minimum brightness states of polarization degree and position angle, with their respective dates. The polarization degree and position angle panels in Figure \ref{optical_all} exhibit random and significant variations around the average values.

\section{Polarimetric variability Analysis}

Variability at long and short timescales is related with the radiative processes that occur in the emitting region and to the strength of the magnetic field.  In the framework of the synchrotron radiation,  the Fermi-accelerated electrons described by a power law with spectral index $\alpha_e$  are injected into the emitting region.  This electron population confined by a randomly oriented magnetic field radiates a photon distribution with a spectral index $(\alpha_e-1)/2$. The size of the emitting region can be constrained using the variability timescale $\tau_{v}$ through the following relation

\be\label{rd}
r_{d} =  \frac{\delta_D}{(1+z)}\tau_{v,min}\, \leq \,\frac{\delta_D}{(1+z)}\tau_{v},
\ee
and the magnetic field can be estimated through the synchrotron cooling timescale 
\be
\small{t}_{\rm syn}=\frac{6\pi}{\sigma_T}\sqrt{\frac{m_e\,q_e\,(1+z)}{\delta_D\,\epsilon_R\,B^3}}
\ee
with $\tau_{v}\simeq t_{\rm syn}$, and then it can be written as 
\be
B=\left(\frac{ 36\pi^2\,m_e\,q_e\, (1+z)}{\sigma^2_T\, \epsilon_R \tau^2_{v}} \right)^{1/3}\, \delta_D^{-1/3}\,.
\ee

Here, $z$ is the redshift, $m_e$ is the electron mass, $\sigma_T$ is the Thompson cross section, $q_e$ is the electric charge,  $\delta_D$ is the Doppler factor and $\epsilon_R$ is the energy of  the R-band associated with the effective wavelength of $\lambda=6400$ \AA \citep[see, ][]{1999PASP..111.1223N,2013ApJS..206...11S}.   It is worth noting that the evolution of the magnetic field structure can be found following \cite{1962SvA.....5..678K} and \cite{1985MNRAS.216..241B}.  \citeauthor{1985MNRAS.216..241B} considered that one or more components could be present, and then each of these could dominate the emission at different epochs.   Variations in the polarization degree $P$ are directly related with the evolution of the magnetic field structure in the emitting region, where the resulting magnetic field  is produced by an ordered ($B_0$; produced by shocks) and a chaotic ($B_c$; immersed in the emitting region) magnetic field.   Considering the spectral power index of the photon distribution radiated in this process, at the limit of $\beta=B_0/B_c\ll 1$, the polarization degree and electron power index are related by  

\be\label{pvsalpha}
P=\frac{(\alpha_e+3)(\alpha_e+5)}{32}\Pi_0\beta^2\,,
\ee

where $\Pi_0=(\alpha_e+1)/(\alpha_e+7/3)$ is the polarization degree of a perfectly uniform magnetic field.\\
\\
In particular, the timescale  for flux variability can be estimated through the relation \citep{1974ApJ...193...43B}
\begin{equation}\label{var}
\tau_{\nu}= \mid \frac{d \ln F}{dt}  \mid^{-1}\,,
\end{equation}

where $dt$ is the time interval between two consecutive fluxes $F_i$ and $F_j$.  For a data set with  $i, j=1,...,N-1$ with $N$ the number of observations, the minimum timescale can be found  by
 
\begin{equation}\label{variab}
 \tau_{\rm v, min}=\mbox{min}\{\tau_{\nu}\}\,. 
\end{equation}

The minimum timescale  equation is used in subsection 5.5.   For our statistical analysis, it is important to define the amplitude of the variations $Y$(\%), fluctuation index  $\mu$,  the fractional variability index $\cal F$ which are \citep{1996A&A...305...42H}
 
\begin{equation}
Y(\%) = \frac{100}{\cal h S i}\sqrt{(S_{\rm max}-S_{\rm min})^2-2\sigma^2_c} \;\; ,
\end{equation}

\begin{equation}
\mu = 100\frac{\sigma_S}{\cal hSi}\% \; ,
\end{equation}

and

\begin{equation}
{\cal F} = \frac{S_{\rm max}-S_{\rm min}}{S_{\rm max}+S_{\rm min}} \;,
\end{equation}

respectively.   Here $S_{\rm i}$ for i=max and min are  the optical flux, the polarization degree and the position angle,  ${\cal hSi}$ is the average value, and $\sigma^2_c =\sigma^2_{\rm max}+\sigma^2_{\rm min}$.  

The long-term R-band photopolarimetric data correspond to 8 year of observations of Mrk\,421. During the monitoring period the blazar displayed several flares, in particular a set of maxima values were observed. In brightness, a value of R$=11.29\pm 0.03$ mag ($93.60\pm  1.53$ mJy) was detected in 2013 April 10, this is our maximum brightness detection. A month later, a maximum for the polarization degree of P=$11.00\pm 0.18\%$ occurred in 2013 May 13. Finally, a maximum for the position angle $\theta=285.4^\circ\pm 8.1$ was detected in  2014 March 27 (see Figure \ref{optical_all}). Considering the entire data, Mrk\,421 exhibited an average polarization degree of $\sim$ 3.6\% with a preferred position angle direction of $\sim169^\circ$.  The low average value of  the polarization degree and the preferred position angle direction are explained in terms of two different emitting regions where the dominated optical flux comes from a region permeated by a stable magnetic field \citep{1983Afz....19..199H, 2016MNRAS.462.4267J}.

In order to look for correlations per year between the optical flux, polarization degree, position angle and normalized Stokes parameters $q$ and $u$\footnote{The normalized Stokes parameters are computed as  {\small $q=Q/F_R=P\cos2\theta$} and {\small $u=U/F_R=P\sin2\theta$}.}, the Pearson's correlation coefficients\footnote{To determine the correlation of the data, we use the Pearson's correlation coefficients which are the covariance of the two variables divided by the product of their standard deviation.} and a statisical analyisis were done and the results are reported in Tables \ref{stat_year} and \ref{pearson_year}, respectively.   
%Table  \ref{stat_year} shows the results of the statistical analysis performed per year.  Ya se dijo arriba.
Among the most outstanding results are: i) In 2012 the object presented the lowest average values in polarization degree and position angle, being equal to $2.60\pm0.05$ \% and  $65.65^\circ\pm0.58$, respectively, ii) in 2013, the maximum variations in the R-band optical flux, polarization degree and position angle were found. The optical flux reached a value of  $65.90$ mJy in a timescale of $\Delta$t= 2 months,   polarization degree went up to $10.65$ \%  in $\Delta$t= 2 months and finally, the position angle increased up to $186.8^\circ$ in $\Delta$t= 4 months, and iii) in 2015 the object exhibited the lowest average flux of  $17.70\pm 0.12$ mJy  with the smallest variations of  $4.96\pm 0.61$ mJy.  It is worth noting that during this year, the highest average polarization degree $6.52\pm0.09$ \% was found. 
%The Pearson's correlation coefficients shown in Table \ref{pearson_year} are computed per year. Ya se dijo arriba.

Also, it is found that the Stokes parameters are strongly correlated in 2010 and 2016, moderately correlated in  2011 and 2013 (anti-correlated in 2014 and 2015) and weakly correlated in 2012 (anti-correlated in 2008 and 2009).

%In this table...
\vspace{1cm}
\section{Multiwavelength data}

To analyze the R-band data in a multiwavelength context, quasi-simultaneous observations from  radio to TeV $\gamma$-ray bands were used. The complete data set used covers the outstanding flaring events of Mrk\,421 that occurred in May 2008, March 2010, July/September 2012 and April 2013.

\paragraph{TeV and GeV observations.}  VERITAS  observed Mrk\,421 in 2007 - 2008 and collected data during 47.3 hours. Details on  data reduction can be consulted in \cite{2011ApJ...738...25A}.  The Whipple 10m telescope observed this source with the ON/OFF and TRK (tracking) mode. The data set in the year 2010 accounts for a total of 36 hours. Details about the analysis performed and normalization  of the photon fluxes can be found in \cite{2015A&A...578A..22A}.    MAGIC telescope system performed 11 observations during the flaring state of March 2010. MAGIC collected data during 4.7 hours of this flare.  Details on the weather conditions, technical problems and analysis can be found in \cite{2015A&A...578A..22A}.  With a total exposure of 49 hours   this object was observed for 10 consecutive nights (from 2013 April 10 to 19) during the flare of April 2013\footnote{http://tdx.cat/handle/10803/290265}.  The ARGO-YBJ data used in this work are reported in \cite{2016ApJS..222....6B}.  The GeV $\gamma$-ray fluxes  were obtained between the energy range 0.1 - 300 GeV using the public database of Fermi-LAT\footnote{http://fermi.gsfc.nasa.gov/ssc/data}.  The reduction of these data was obtained following \cite{2013MNRAS.434L...6C}.

\paragraph{X-ray observations}.   The Swift-BAT/XRT  data used in this work are  publicly available \footnote{http://swift.gsfc.nasa.gov/cgi-bin/sdc/ql?}.  The XRTE - PCA (ASM) data were obtained from the public MIT archive\footnote{http://heasarc.gsfc.nasa.gov/docs/xte/asm\_products.html}.   The MAXI data are publicly available\footnote{http://maxi.riken.jp/top}. 

\paragraph{Additional optical observations}.  Several campaigns were performed with the GLAST-AGILE Support Program (GASP)  within the Whole Earth Blazar Telescope (WEBT)\footnote{http://www.oato.inaf.it/blazars/webt/},  and New Mexico Skies, Rovor,  Blandford and the Perkins  telescopes.  Optical polarization measurements are shown from the Crimean and St. Petersburg observatories.  The Swift - UVOT observations are given in the three bands, UVW1, UVM2, and UVW2 which are available at the HEASARC data archive\footnote{http://swift.gsfc.nasa.gov/cgi-bin/sdc/ql?}.

\paragraph{Radio observations}. The Mets\"ahovi radio telescope measurements were made at 36.8 GHz. The description of the data reduction and analysis is given in  \citet{1998A&AS..132..305T}. UMRAO provided data at 4.8 GHz, 8 GHz and 14.5 GHz between June 2006 and May 2008. The calibration and reduction procedures have been described in \citet{1985ApJS...59..513A} \footnote{http://www.physics.purdue.edu/astro/MOJAVE/sourcepages/1101+384.shtml}.   OVRO \citep{2011ApJS..194...29R}  data are publicly available\footnote{http://www.astro.caltech.edu/ovroblazars/}.   Carma, as a part of the MARMOT\footnote{http://www.astro.caltech.edu/marmot/} blazar monitoring project, has performed  observations since February 2013 using the eight 3.5 m telescopes of the array with a central frequency of 95 GHz.

\section{Multiwavelength data Analysis}

In order to analyse the entire multiwavelength data collected for Mrk\,421, these have been divided per year. In the next subsections, the most important results of the analysis of the variability behavior around the observed flares are presented.

\subsection{Flaring event in May 2008}

\begin{figure*}
\centering
\includegraphics[width=0.9\textwidth]{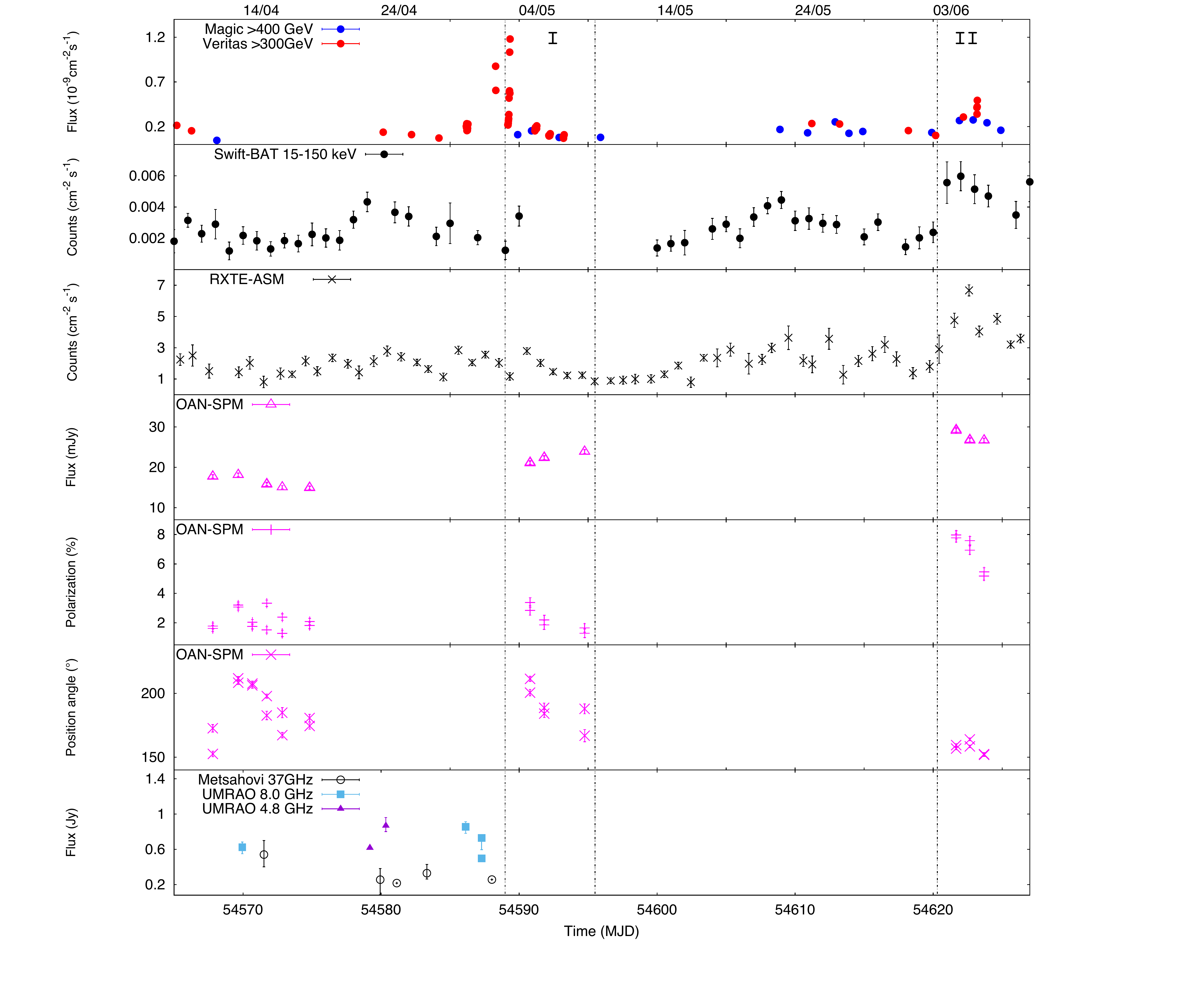}
\caption{Mrk\,421 lightcurves including polarization and position angle data between 2008 April 22 (MJD 54578) and June 08 (MJD 54625) collected with Veritas, Swift, RXTE, OAN-SPM, GASP-WEBT, Mets\"ahovi and UMRAO are shown.  From top to bottom: TeV $\gamma$-rays, hard/soft X-rays, optical flux, polarization degree, position angle and radio flux variations are presented. Dashed vertical lines indicate the two analyzed periods.}
\label{flare_2008}
\end{figure*}

\begin{figure*}
\centering
\includegraphics[width=\textwidth]{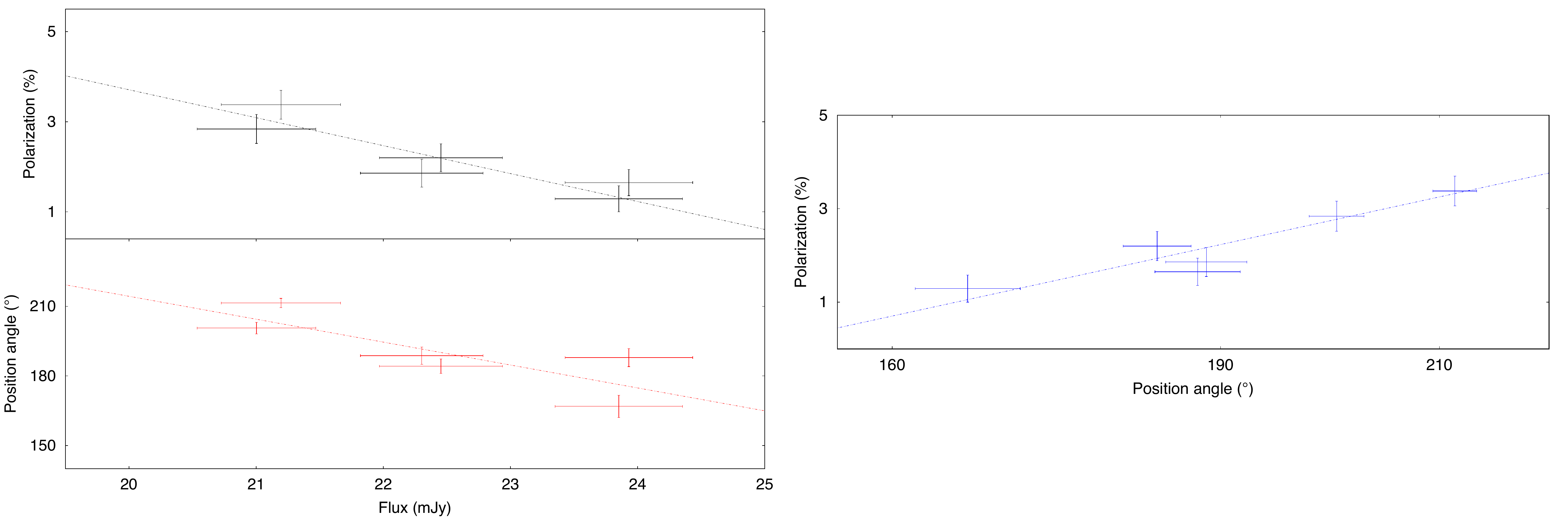}
\caption{R-band photometric correlations found between 2008 May 04 (MJD 54590) and 2008 May 08 are shown. Left: Optical flux vs  polarization degree (top panel) and optical flux vs  position angle (bottom panel) correlations are shown. Right: Polarization degree vs position angle correlation is presented. The Pearson's correlation coefficients are given in Table \ref{pearson_all} (column 2).}
\label{correlation_2008}
\end{figure*}

The R-band optical light curve together with polarization degree and position angle variability presented in Figure \ref{flare_2008} shows the variability behavior of Mrk\,421 from 2008 April 08 (MJD 54564) to 2008 June 08 (MJD 54620). This figure also shows the TeV $\gamma$-ray,  hard/soft X-ray and radio observed variations.  During the first period from 2008 May 04 (MJD 54590) to 08, a bright TeV $\gamma$-ray flux was observed  (reaching the level of 10 Crabs) that was not detected in soft/hard X-ray activity, suggesting a possible ``orphan flare''  \citep{2011ApJ...738...25A,2015APh....71....1F}.  The TeV flare coincides with a significant drop of 60$^\circ$ in the position angle and with a polarization degree of  2\%. Subsequently, the TeV high-flux level decreased until it reached the quiescent level whereas the optical flux continued to increase gradually. The Pearson's correlation values are reported in Table \ref{pearson_all} (column 2). These values were estimated considering a maximum allowed time difference between data of  $\Delta t\leq$ 0.45 days \citep{2011ApJ...738...25A}.  The analysis performed during the period in which the TeV $\gamma$-ray flux began to decline, revealed that optical flux is anti-correlated with TeV $\gamma$-rays, hard/soft X-rays, polarization degree and position angle.  Similarly, this analysis shows that the polarization degree and the position angle are strongly correlated.  Figure \ref{correlation_2008} shows the correlation between polarization degree and position angle (right panel), and the anti-correlations of the optical flux with the polarization degree and  the position angle (left panels). The previous result indicates that there is a TeV $\gamma$-ray emitting region with a high ordered magnetic field that is different from the zone where the optical flux is released \citep{2008Natur.452..966M}.

%During the period between 2008 March 14 (MJD 54590) and 2008 June 03 (MJD 54620), there was lack of TeV $\gamma$-ray data. However, an evident flare occurred in X-ray and optical bands. 
Recently, \cite{2016A&A...593A..91A} reported a TeV $\gamma$-ray flare during the second period (from 2008 June 03  to 08). With our data, we found that this TeV flare is strongly correlated  with the polarization degree percentage (with a Pearson's coefficient of 0.88 and p-value of 0.02), and without any significant change of the position angle. Assuming this hypothesis is valid, a possible mechanism to explain this variability behavior is by means of perpendicular shocks moving along a uniform-axially symmetric jet axis.  In these cases, the jet plasma is compressed, and then the degree of polarization varies without  significantly altering the value of the position angle \citep{2010Natur.463..919A, 2008Natur.452..966M}.  In addition, using equations \ref{var} and \ref{variab},  the optical flux varies on timescales of 15.4 days and  the minimum variability timescale is $\sim$ 9.6 hours.  Both results are similar to those found in \cite{2011ApJ...738...25A}.\\

\subsection{Flaring event in March 2010} 

\begin{figure*}
\centering
\includegraphics[width=0.95\textwidth]{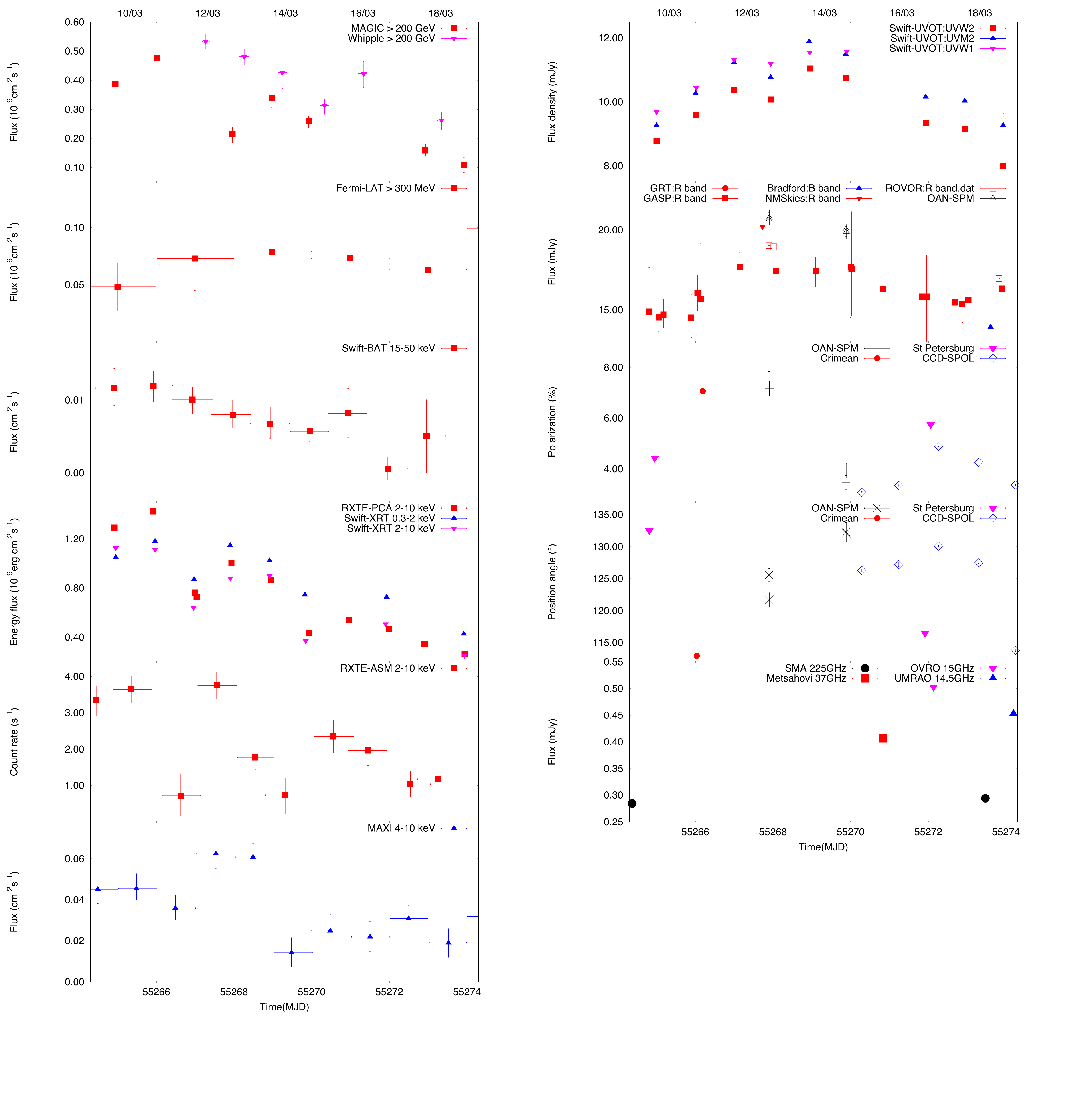}
\caption{Mrk\,421 lightcurves are shown including polarization and position angle data between 2010 March 09 to 19 obtained with multiple satellites and ground based observatories. From top to bottom: (Left) TeV $\gamma$-rays, GeV $\gamma$-rays and hard/soft X-ray variations are presented. (Right) UV, optical R-band fluxes,  polarization degree, position angle and radio wavelengths variations are presented.}
\label{flare_2010}
\end{figure*}

Figure  \ref{flare_2010} shows the R-band photopolarimetric observations along with TeV $\gamma$-ray, hard/soft X-ray, optical and radio data. Although there are only four photopolarimetric data points, 2 obtained in 2010 March 12 (MJD 55267) and two in 2010 March 14 (MJD 55269), these points (black empty triangle) are useful because between 2010 March 10 (MJD 55265) to 22 (MJD  55277) Mrk\,421 exhibited high activity level in all bands. Considering R-band photopolarimetric data and the maximum allowed time difference {\rm $\Delta t\leq$ 0.45 days}, the values of the Pearson's correlation coefficients and p-values were computed and reported in Table \ref{pearson_all} (column 3).  But it is worth to remember that these correlations are based on four points only.  Based on this analysis, the optical flux is found to be correlated with TeV $\gamma$-rays, hard/soft X-rays and the polarization degree, and anti-correlated with the position angle and the Stokes parameters.  During this flare, the polarization degree and the position angle exhibited variations, 4\% in the polarization degree and 10$^\circ$ in the position angle.

\subsection{Flaring event in the year 2012}

\begin{figure*}
\centering
\includegraphics[width=0.95\textwidth]{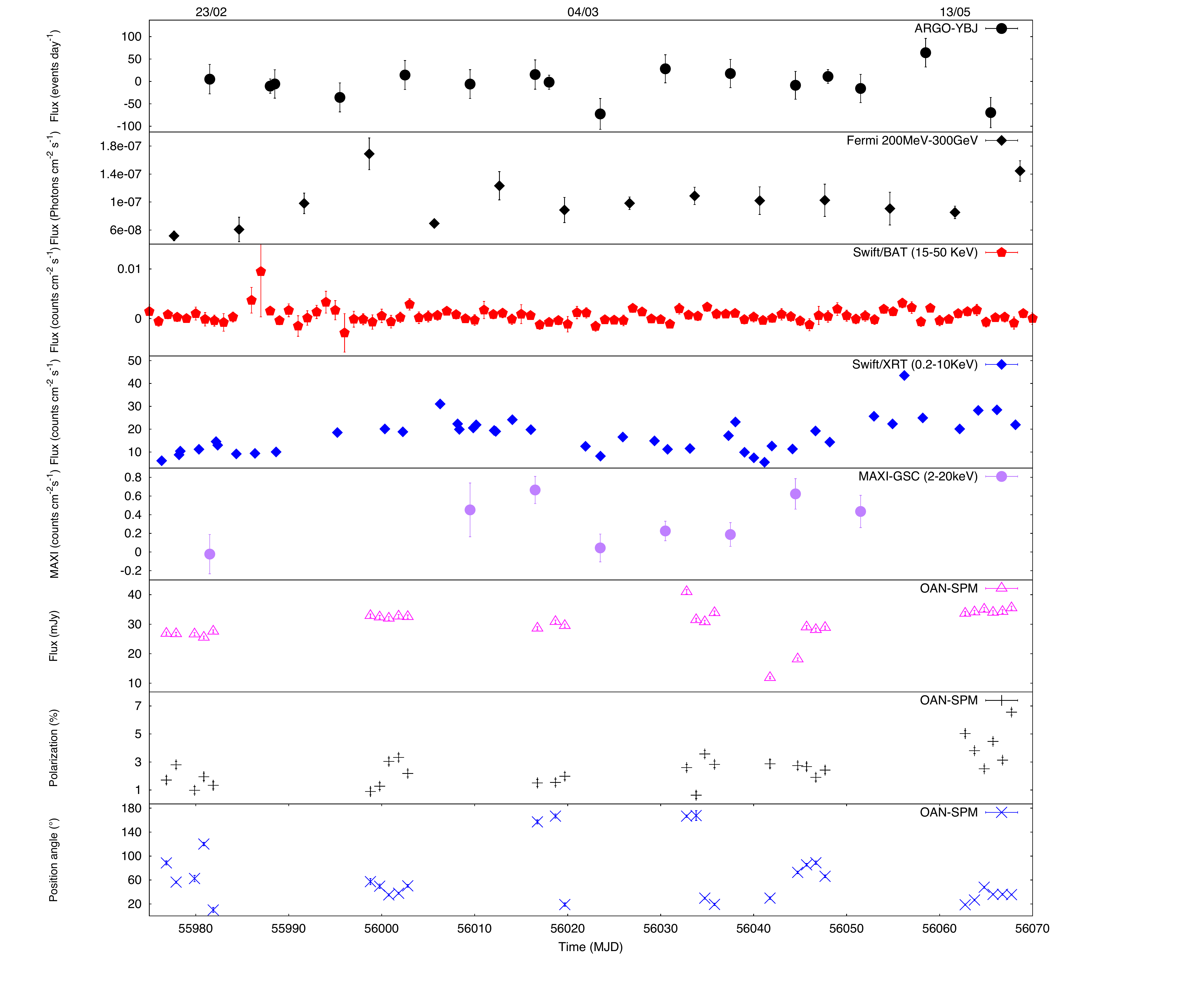}
\caption{Mrk\,421 lightcurves including polarization and position angle data between 2012 February 18 and May 23 obtained with ARGO-YBJ, Fermi, Swift, MAXI-GSC and OAN-SPM are shown. From top to bottom: TeV $\gamma$-rays,  GeV $\gamma$-rays,  hard/soft X-rays,  optical flux, polarization degree and position angle variations are presented.}
\label{flare_2012}
\end{figure*}

R-band photopolarimetric observations complemented with TeV $\gamma$-ray, hard/soft X-ray, and radio lightcurves are shown in Figure  \ref{flare_2012}.  The lightcurves include the period from 2012 February 19 (MJD 55976) to 2012 May 20 (MJD 56067).  During this period, Mrk\,421 shows a moderate activity in some wavebands (e.g.  the optical flux varied from 40.96 to 11.81 mJy in $\Delta$t =9 days when the polarization degree and the position angle showed variations of $\sim 0.3\%$ and $\sim 137^\circ$, respectively).  Afterwards, this source went into a long-lasting outburst phase starting in 2012 July 9 (MJD 56117) and ending in 2012 September 17 (MJD 56187) when this object was not visible with optical telescopes.  Only some instruments, including ARGO-YBJ \citep{2012ATel.4272....1B}, Fermi  \citep{2012ATel.4261....1D}, BAT, NuSTAR \citep{2016ApJ...819..156B} and OVRO \citep{2012ATel.4451....1H} could observe Mrk\,421 outburst. Therefore, Fermi collaboration reported  the highest flux observed by this source since the beginning of the Fermi mission \citep{2012ATel.4261....1D} and ARGO-YBJ also detected a high-flux level  \citep{2012ATel.4272....1B}. Although  this flare was intensively followed  by Fermi-LAT and ARGO-YBJ,  MAXI-GSC and Swift-BAT instruments did not report the same high activity in X-ray bands.

\subsection{Flaring event in April 2013}

\begin{figure*}
\centering
\includegraphics[width=0.95\textwidth]{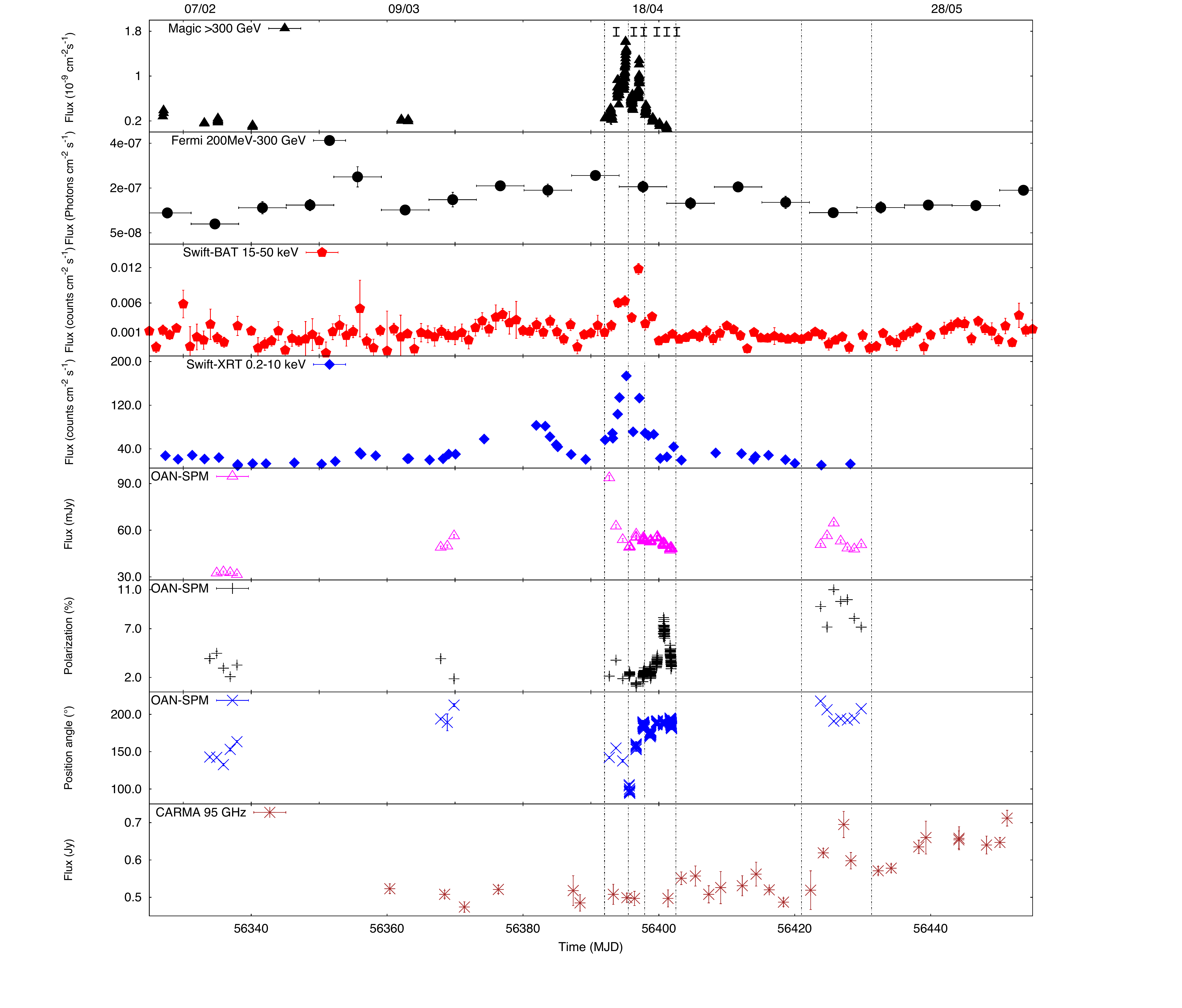}
\caption{Mrk\,421 lightcurves including polarization and position angle data between 2013 January 28 and June 17 obtained with Magic, Fermi, Swift, OAN-SPM and CARMA are shown. From top to bottom:  TeV $\gamma$-rays, GeV $\gamma$-rays,  hard/soft X-ray,  optical flux,  polarization degree, position angle and radio fluxes variations are presented. April data were divided in three periods, shown with vertical dashed lines. May data are shown within vertical dashed lines.}
\label{flare_2013}
\end{figure*}

\begin{figure*}
\centering
\includegraphics[width=0.95\textwidth]{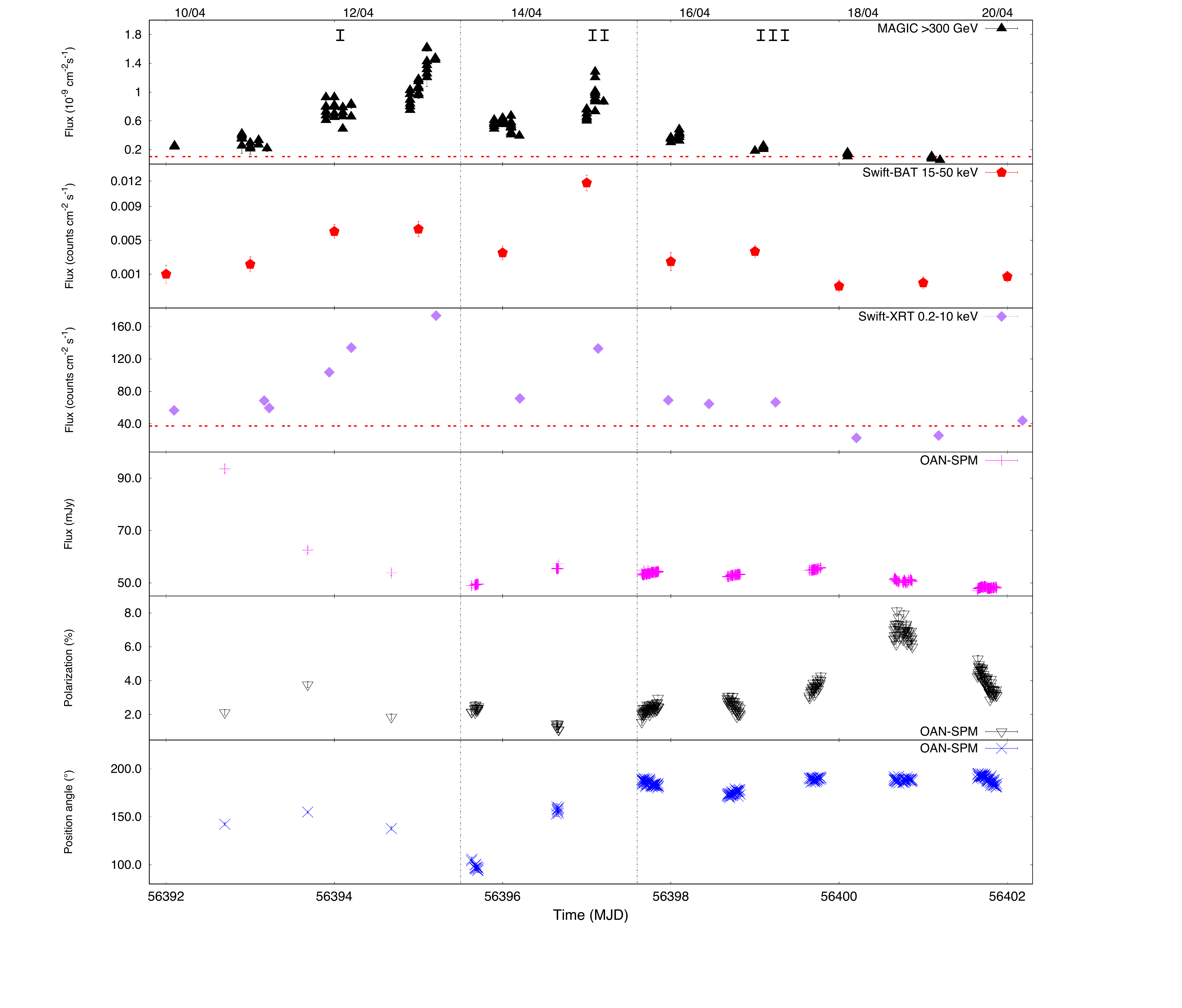}
\caption{A blow-up of the ten-day high activity observed in April 2013 (for details see caption in Figure \ref{flare_2013}).  The vertical dashed lines separate three periods: The first period from April 10 to 13 shows the strongest activity, the second period from April 13 to 15 displays the intermediate level activity  and the third period from April 15 to 19 exhibits slowly decreasing fluxes in all bands.}
\label{flare_8days}
\end{figure*}

The multiwavelength lightcurves including polarization degree and position angle variations from 2013 January 02 (MJD 56300) to 2013 July 27 (MJD 56500) are presented in Figure \ref{flare_2013}.  This strong flare was widely monitored in several energy bands and was observed from 2013 April 09 (MJD 56391) to 19 in TeV \citep{2013ATel.4976....1C}, GeV \citep{2013ATel.4977....1P}, hard/soft X-ray \citep{2014A&A...570A..77P, 2013ATel.4974....1B,2013ATel.4978....1N} and optical \citep{2013ATel.4982....1S, 2015arXiv150801438F} bands.   A blow-up of the lightcurves around the flare are shown in Figure \ref{flare_8days}.   The R-band optical light curve together with polarization degree and position angle shown in this figure includes 235 observations which were obtained during 10 consecutive nights.  The photopolarimetric data were divided in three periods in accordance with the activity state of the source.  These periods were defined considering the TeV and X-ray data. The lowest activity state was defined taking into account the fluxes observed in April 18 and 19 in both bands.  This is shown with red-dashed lines in Figure \ref{flare_8days}. These lines indicate the  average fluxes $(0.11\pm 0.04)\times 10^{-9}\,{\rm cm^{-2}\,s^{-1}}$  and $37.26\pm 0.04\, {\rm counts\,cm^{-2}\,s^{-1} }$ for  TeV and X-ray bands, respectively.  Therefore, high activity states are defined when data are 3$\sigma$ above the red-dashed lines: First period marks the high activity level, the second one a moderate activity level and finally the third period shows the decreasing activity level.\\

\paragraph{\sf First period: From 2013 April 10  (MJD 56392) to 13.}

The strongest flare was observed.  The TeV $\gamma$-ray and optical fluxes were the highest ever recorded for this object \citep{2013ATel.4976....1C}.  In the optical R-band, the flux reached an unprecedented value of 93.60$\pm$1.53 mJy (11.29$\pm$0.03 mag), which corresponds to an optical luminosity of $L_{\rm R}\simeq9.3 \times 10^{44}$ erg/s.  The results of the statistical analysis of these data are presented in Table \ref{episodes}.  It can be seen that while a difference in flux level of $\Delta$F=39.78 mJy was registered, no significant variations were found in the polarization degree and the position angle, i.e. these parameters varied  $\sim$1\% and $\sim$ 10$^\circ$ over their average values, respectively. 

Column (4) in Table \ref{pearson_all} shows the Pearson's correlation coefficients among TeV $\gamma$-ray, hard/soft X-ray, optical and radio  fluxes, the polarization degree, the position angle and the Stokes parameters by considering the maximum allowed time difference $\Delta t\leq$ 0.45 days.  The analysis performed indicates that  optical R-band is strongly anti-correlated with TeV $\gamma$-rays, hard/soft X-ray bands, and with the Stokes parameter $q$, and correlated with the radio wavelength, although high activity was not detected in this band.    A strong correlation between the polarization degree and the position angle was found and a moderate correlation was found between the Stokes parameters.

\paragraph{\sf Second period: From 2013 April 13 (MJD 56395) to 15.}

During this period, the object showed a moderate activity level. The statistical analysis is also reported in Table \ref{episodes} and  whereas no significant variations were found in the polarization degree since it varied only  $\sim$ 1\% over its average value, the position angle displays a long rotation 
of $\sim$100$^\circ$. Column (5) in Table \ref{pearson_all} shows the Pearson's correlation coefficients among TeV $\gamma$-ray, hard/soft X-ray, optical and radio  fluxes, the polarization degree, the position angle and the Stokes parameters for a maximum allowed time difference of $\Delta t\leq$ 0.45 days. The analysis indicates that the optical R-band is strongly anti-correlated with soft X-ray band and correlated with hard X-ray band as well as the radio waveband (GHz).  In addition, optical flux is correlated with the position angle and the Stokes parameter $U$, and the polarization degree is anti-correlated with the optical flux and the position angle.  Their respective Pearson's coefficient values are reported in Table \ref{pearson_all} (column 4).  

\paragraph{\sf Third period: From April 15 (MJD 56397) to 19.}

From the multiwavelength lightcurves (Figure \ref{flare_8days}) it can be seen that TeV $\gamma$-ray, hard/soft X-ray and  optical fluxes show a tendency of decreasing activity level.   No correlation was found among fluxes, the polarization degree, the position angle and Stokes parameters (see  column 6 in Table \ref{pearson_all}).  Figure \ref{flare_8days} shows that while the value of the position angle had very small variations  around $\sim$ 185$^\circ$, the polarization degree increased substantially, reaching the value  of 8.33\% around 2013 April 18 (MJD 56400). After this value, it started decreasing again.

\subsection{R-band intranight polarimetric variability observations}

\begin{figure*}
\centering
\includegraphics[width=0.85\textwidth]{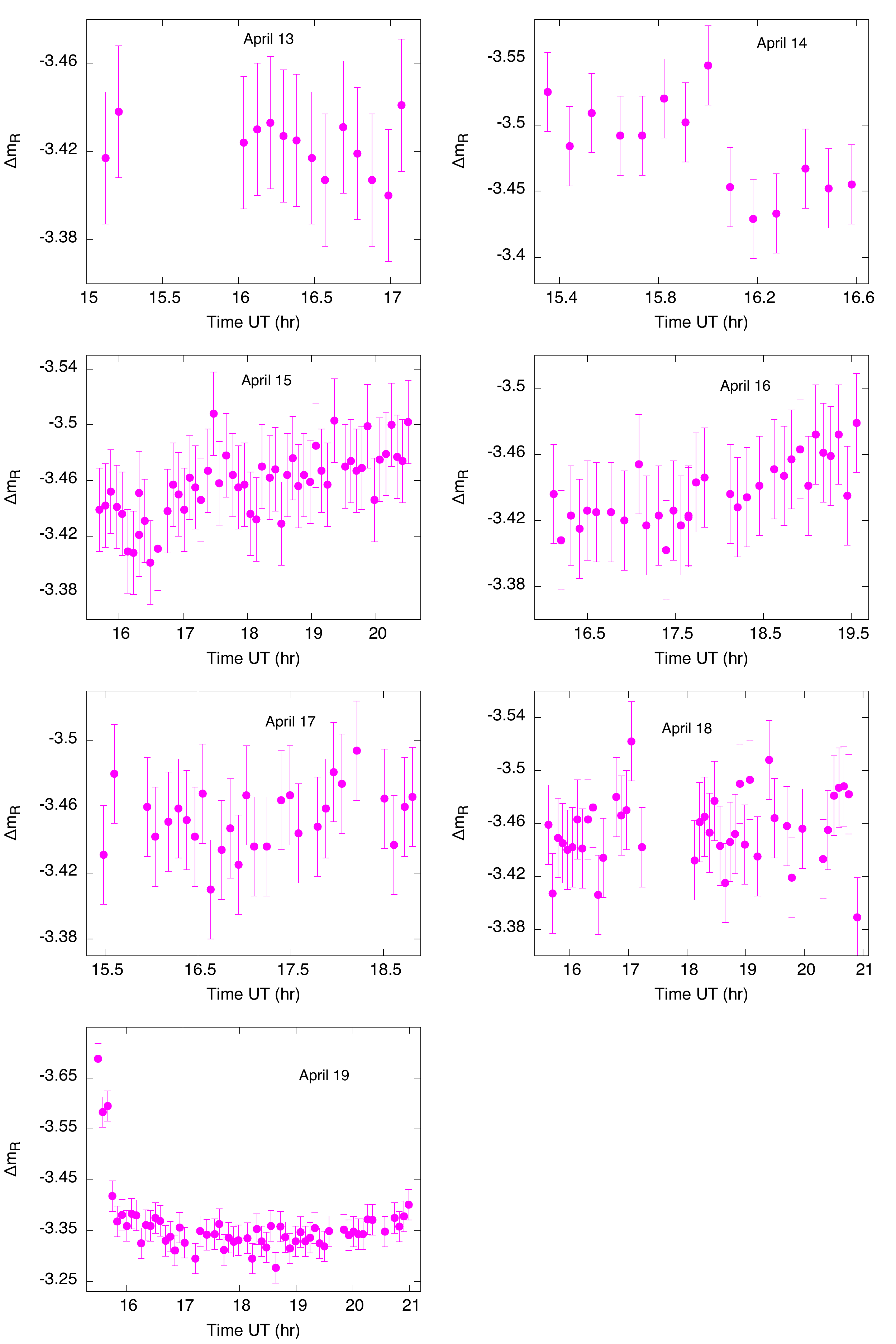}
\caption{ R-band differential photometry data from 2013 April 13 to 19. Each of the seven panels show 
the changes in the differential R-band magnitudes. The duration of the monitoring in hours from the beginning to the end for each night is shown in the X axis.}
\label{Intraphot}
\end{figure*}

\begin{figure*}
\centering
\includegraphics[width=0.85\textwidth]{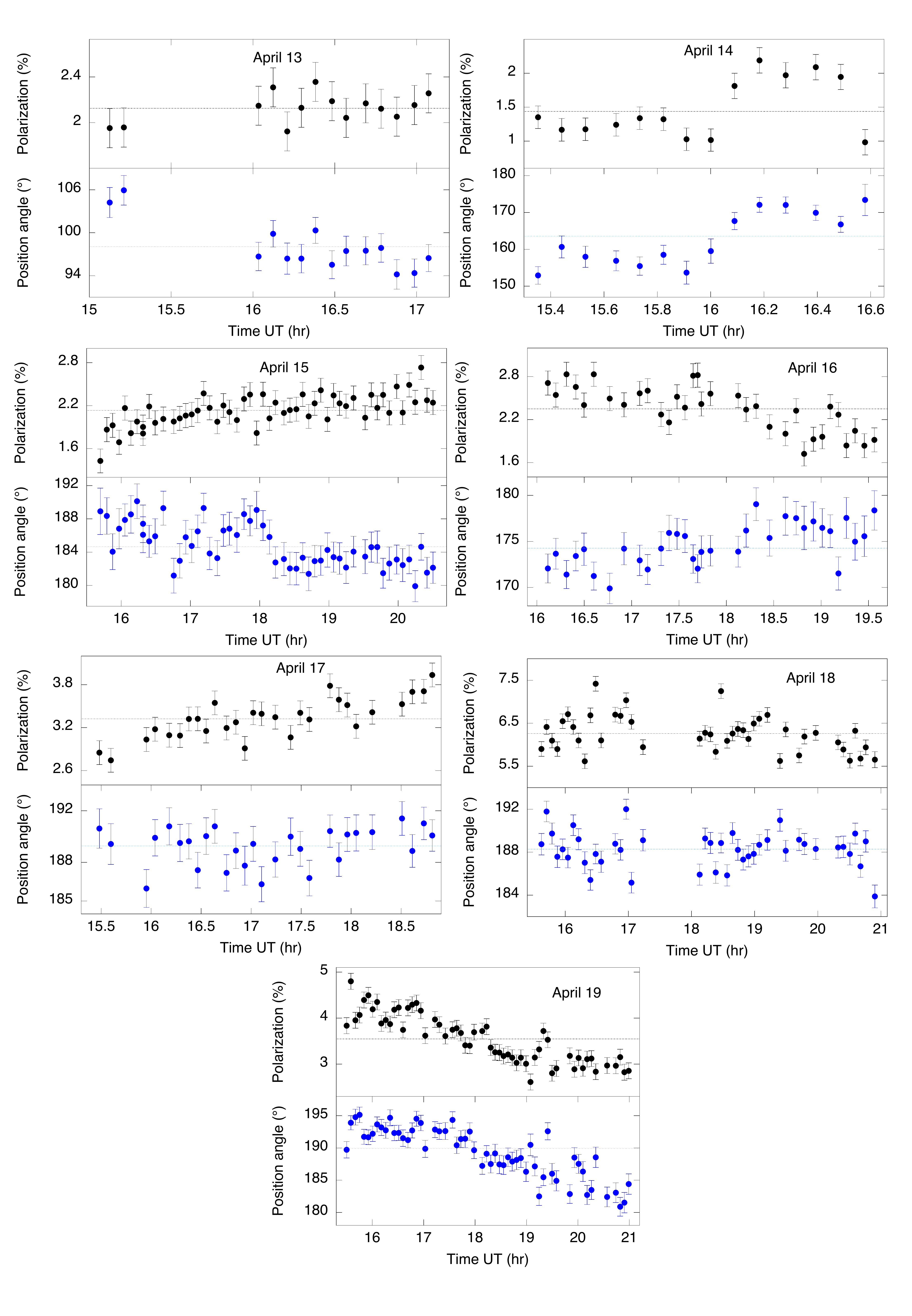}
\caption{ The intranight polarimetric data from  2013 April 13 to 19. Each of the seven panels show the  polarization degree (top) and the position angle (bottom) variations. 
The duration of the monitoring in hours from the beginning to the end for each night is shown in the X axis.}
\label{Intrapol}
\end{figure*}

Due to the high activity level observed during April 2013 in Mrk\,421  (first period), a photopolarimetric monitoring campaign dedicated to detect intranight variations started three days after the maximum flux observed in the R-band data.  This monitoring has a cadence of $\sim$ one day. It was performed during 7 consecutive nights with a duration that ranges from 1 to 5 hours. The intranight variability observations were done from 2013 April 13 to 19 at the OAN-SPM and includes 14, 14, 53, 35, 28, 33 and 58 observational points per night, respectively. In Figure \ref{Intraphot}, the results of the R-band differential photometry $\Delta\,m_{R}$ are shown. Also, in Figure \ref{Intrapol} the corresponding polarimetric intranight variations are shown. In these figures there are clear hints of possible intranight variability in some nights. To confirm this, a statistical analysis of the data was done using the {\bf F} and ANOVA tests and following \citet{2010AJ....139.1269D, 2014AJ....148...93D}. On the one hand, for the {\rm F}-test a star with a magnitude that is only 0.08 magnitudes brighter than Mrk\,421 was used, i.e. both objects have virtually the same photometric errors. On the other hand,  ANOVA compares the dispersion between different data groups drawn out from the light curve of a single object, rather than comparing between different lightcurves of different objects. The detection limit of intranight variability was set at a level of significance $\alpha=10^{-3}$, i.e. a probability of 0.1\% that the light curve is obtained from non-variable sources.  However, a significance $\alpha=10^{-2}$ is considered as a hint of possible variations. The values obtained and derived from the variability statistical analysis are reported in Table \ref{daily_var} where the p value can be found in column 9.  Hence, using the {\rm F} and ANOVA tests, intranight variability was found in the R-band differential magnitudes and also in polarimetry (degree and position angle) during four nights (2013 April 14, 15, 16 and 19). Intranight variations are a valuable tool for estimate the minimum variability timescale per night using the photometric data. In column (10) the minimum variability timescale per night is given, and the minimum value was found on April 15 where it has a value of 2.34$\pm$0.12 hrs.

\begin{figure*}
\centering
\includegraphics[width=0.65\textwidth]{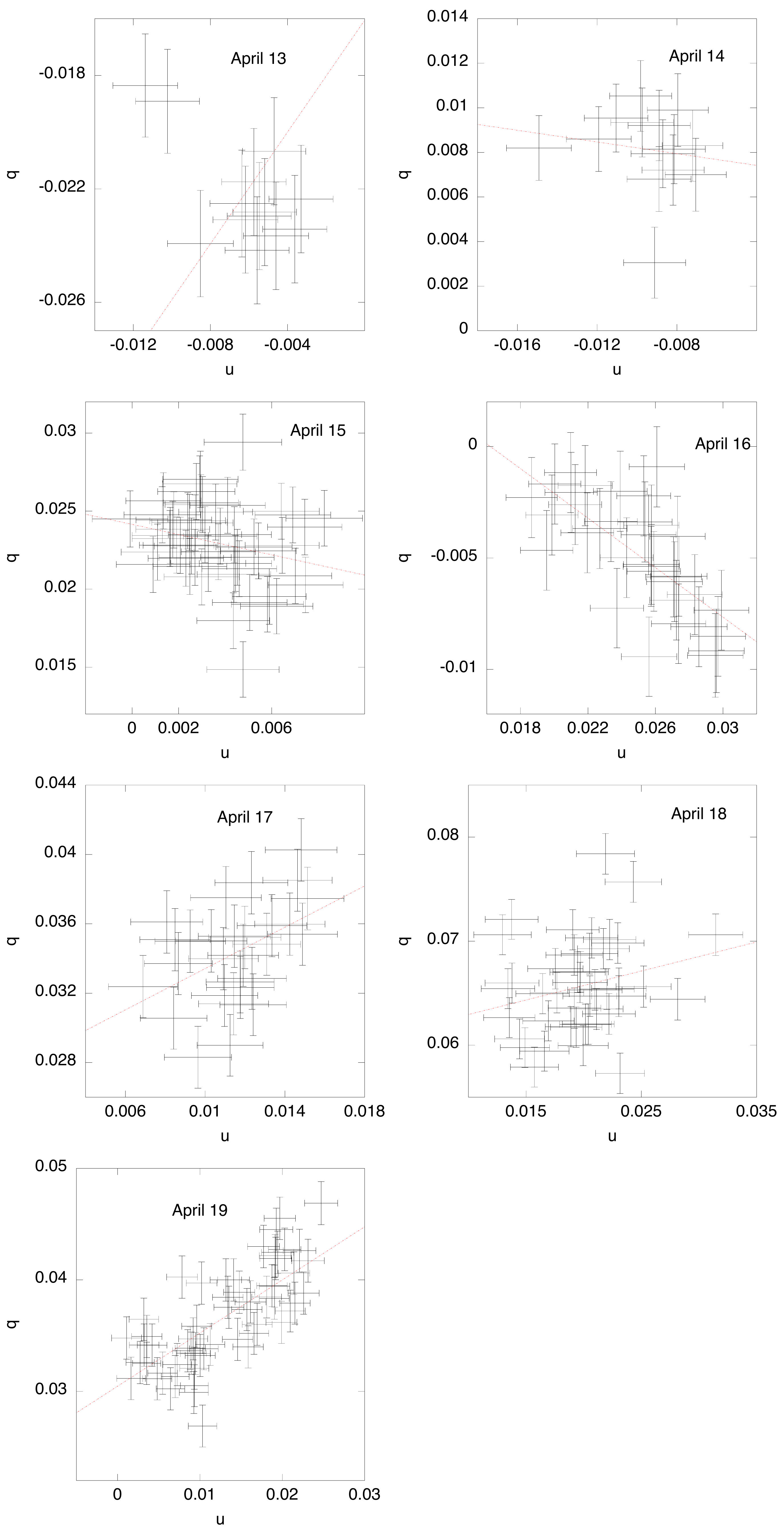}
\caption{Normalized Stokes parameters correlations found in the u-q plane from 2013 April 13 to 19.}
\label{Night_UvsQ}
\end{figure*}

In order to search for correlations between variations in flux, the polarization degree and the position angle, the Stokes parameters were plotted, as shown in Figure \ref{Night_UvsQ}.  The values of the slopes obtained from the fitting and the Pearson's correlation coefficients are reported in Table \ref{nightly_analysis}. 

\begin{figure*}
\centering
\includegraphics[width=1.05\textwidth]{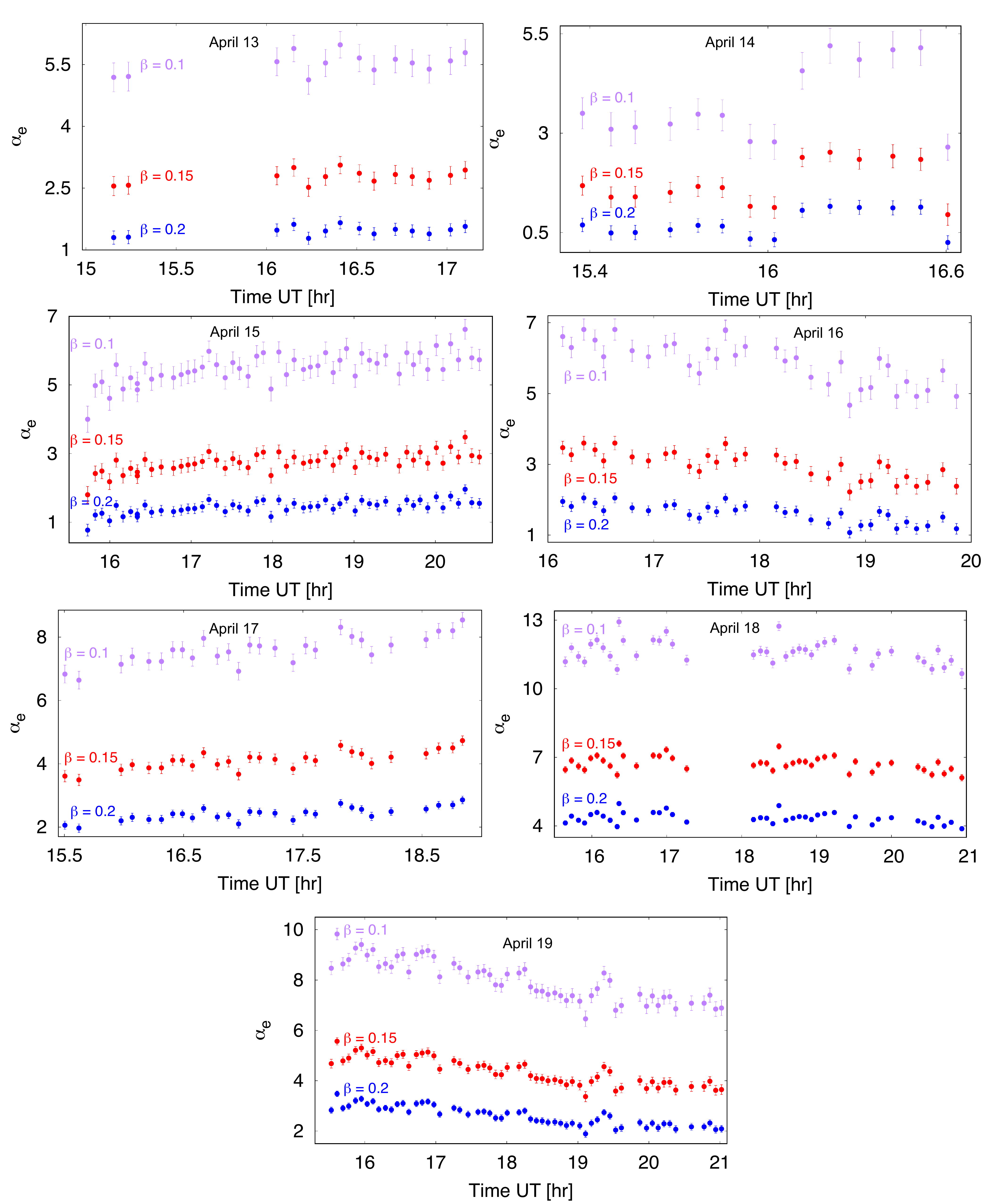}
\caption{The intranight variations of the electron power index as a function of time ($\Delta t\sim$ minutes) and different fractions of ordered and chaotic magnetic fields.}
\label{Spec_index}
\end{figure*}

Figure \ref{Spec_index} displays the intranight variations of the electron power index as a function time ($\Delta t\sim$ minutes) and the different fractions of ordered and chaotic magnetic fields. The electron power indexes were derived from the polarization degree (see eq. \ref{pvsalpha}) during the flaring activity presented between 2013 April 13 (MJD 56395) and 19.  Table \ref{nightly_analysis} shows the average spectral index per night for three values of magnetic field ratios  ($\beta=0.1$, 0.15 and 0.20).  A description per day of the Stokes parameter correlations with the intranight variations of optical flux, the polarization degree and  the position angle is given in the following paragraphs.   Additionally, the maximum changes in the values of spectral index and timescales are reported in Table \ref{nightly_analysis}.

\paragraph{\sf 2013 April 13}. With less than 2 hours of observations during this night, the optical flux, the polarization degree and the position angle display small variations,  0.4 mJy,  0.2\%  and 10$^\circ$, respectively. The Stokes parameters were moderately anti-correlated.  Very fast spectral index variations were detected during this night. The maximum changes in the spectral index were  0.85, 0.54, 0.38 for $\beta$=0.1, 0.15 and 0.20, respectively, on a timescale of  $\sim$ 11 min.

\paragraph{\sf 2013 April 14}. Variations of the optical flux were found during this night.  It varied 57.34 mJy whereas the polarization degree changed only 1.10 \%.    No significant correlation was found between the Stokes parameters. The maximum changes in the spectral index were  0.87, 0.55, 0.37 for $\beta$=0.1, 0.15 and 0.20, respectively, in a timescale of  $\sim$ 39 min.

\paragraph{\sf 2013 April 15}. During this night, the optical flux, the polarization degree and the position angle showed=variations.   No significant correlation was found between the Stokes parameters. The maximum changes in the spectral index were  2.62, 1.68, 1.19 for $\beta$=0.1, 0.15 and 0.20, respectively, in a timescale of  $\sim$ 4.6 hours.

\paragraph{\sf 2013 April 16}. With less than 5 hours of observations during this night, the levels of the optical flux and the position angle increased, whereas the polarization degree decreased. A strong anti-correlation was found between the Stokes parameters. The maximum changes in the spectral index were  2.14, 1.38, 0.9 for $\beta$=0.1, 0.15 and 0.20, respectively, in a timescale of  $\sim$ 2.5 hours.

\paragraph{\sf 2013 April 17}. With almost 3 hours of observations during this night, the levels of the optical flux and the position angle increased slowly whereas the position angle had small changes around its average.  A moderate correlation was found between the Stokes parameters. The maximum changes in the spectral index were  1.90, 1.24, 0.89 for $\beta$=0.1, 0.15 and 0.20, respectively, in a timescale of  $\sim$ 3.2 hours.

\paragraph{\sf 2013 April 18}.  The highest mean polarization value of ($6.86\pm 0.03$)\% was observed, whereas the position angle had small variations. No significant correlation was found between the Stokes parameters. The maximum changes in the spectral index were  2.26, 1.49, 1.10 for $\beta$=0.1, 0.15 and 0.20, respectively in a timescale of $\sim$ 4.6 hours.

\paragraph{\sf 2013 April 19}.  During this night,  the polarization degree and the position angle decreased.  A strong correlation was found between the Stokes parameters. The maximum changes in the spectral index were 3.37, 2.20, 1.59 for $\beta$=0.1, 0.15 and 0.20, respectively, in a timescale of  $\sim$ 3.5 hours.\\

Subsequently, from 2013 May 10 (MJD 56422) to 17 (MJD 56429) a new flare was observed only in the R-band and radio wavebands, as shown in Figure \ref{flare_2013}. Mrk\,421 showed the second brightness state of $64.49\pm1.08$ mJy ($11.70\pm0.04$ mag),  which corresponds  to an optical luminosity of $L_{\rm R}\simeq6.4 \times 10^{44}$ erg/s. Moreover, the highest value of polarization degree of ($11.00\pm0.18$) \% in the R-band was detected. In Table \ref{pearson_all} (column 7) the Pearson's correlation coefficients among fluxes, the polarization degree, the position angle and the Stokes parameters are shown for a maximum allowed time difference of $\Delta t\leq$ 0.45 days.  The values of these coefficients indicate that soft X-ray, optical and radio fluxes are correlated. In the high energy bands no significant correlations were found. It is worth noting that although soft X-ray flux was correlated with the optical and radio wavebands, no high activity was detected. The minimum variability timescale found in this period was $\sim$ 7 days.

\section{Analysis and General Results}

\subsection{Two emitting zones}

A two emitting region model has been proposed for several blazars in order to describe the broadband SED in flaring and quiescent states. Moreover, the superposition of two polarized components (one constant and other variable) coming from different zones has been widely suggested to describe the variability behavior \citep{1984MNRAS.211..497H,1996MNRAS.282..788B, 1993ChA&A..17..229Q}.   Following \cite{1984MNRAS.211..497H},  the resulting polarization degree and the position angle of the two optically thin synchrotron emitting regions can be written as 

\be
P^2=\frac{P^2_{\rm con}+P^2_{\rm var}I^2_{\rm v/c}+2P_{\rm con}\,P_{\rm var}I_{\rm v/c}\cos2\xi}{(1+I_{\rm v/c})^2}
\ee

and

\be
\tan2\theta=\frac{P_{\rm con}\sin2\theta_{\rm con}+P_{\rm var} I_{v/c}\sin2\theta_{\rm var}}{P_{\rm con}\,\cos2\theta_{\rm con}+P_{\rm var}I_{v/c}\cos2\theta_{\rm var}}
\ee

respectively,  where $\xi=\theta_{\rm con} - \theta_{\rm var}$ and $I_{v/c}=F_{\rm var}/F_{\rm con}$ is the  flux ratio of different zones, one variable and one constant. Taking into account the obtained average Stokes parameters $U_c=(1.65\pm0.01)$ mJy and $Q_c=(0.35\pm0.01)$ mJy, and following \cite{1985ApJ...290..627J}, the constant values found are $P_{\rm con}=3.25\pm0.70\%$ and  $\theta_{\rm con}=219.01^\circ\pm0.02$. Considering the limit cases $F_{\rm var}>>F_{\rm con}$ and $F_{\rm var}<<F_{\rm con}$, the resulting polarizations are  $P\simeq  P_{\rm var}$ and $P\simeq  P_{\rm con}$, respectively. 

The previous results suggest that the optical emission observed during the 2013 flares could be produced in two different regions through the injection of additional high-energy electrons.

The first region is related with the flare observed in April 2013 and the second one to the flare detected in May 2013.  The optical flux associated to the emitting region that originates the TeV gamma-rays and the X-rays has a  polarization degree of $P_{\rm con}=3.25\pm0.70 \%$ with small variations and the region associated to the emission observed in optical and radio wavebands has a variable polarization degree.  \cite{2013ApJ...774...18Z} presented a full three-dimensional radiation transfer code, considering a helical magnetic field and geometry effects through the jet.  Synchrotron radiation in this code is assumed to come from an ordered magnetic field and SSC is handled with the two-dimensional Monte-Carlo/Fokker-Planck code.  Using several scenarios for leptonic models in flaring events,  they showed that depending on the energy range analyzed  the blazar Mrk 421 had different behaviors in the polarization degree and position angle. In particular, in the scenario for the injection of additional high-energy electrons, the polarization properties at lower energies had small variations (quasi-constant). At higher energies, the polarization degree was expected to vary significantly. This scenario would show that two different electron populations could describe the flares observed in April 2013 and later in May. The additional high-energy electrons had sufficiently cooled down, in order to that former electron population is dominant again.

It is common to suggest that the low polarization degree observed in Mrk\,421 might be associates to the emission originated from the accretion disc or emitting regions outside of the jet \citep{2016MNRAS.462.4267J}. However, it is shown that both emitting optical fluxes are originated by synchrotron radiation where one flux is much more polarized than the other. 

\subsection{Modeling the Spectral Energy Distribution}

The most widely adopted scenario to explain the  broadband SED of HBL objects with the fewest parameters,  is the one-zone SSC model.    In this context, the electrons within the emitting region are moving at ultra-relativistic velocities in a collimated jet and are confined in this region by the magnetic field.   Then, photons are radiated via synchrotron emission and  up-scattered to higher energies by the the same electron population via inverse Compton scattering. Photons from radio to X-ray bands are interpreted as synchrotron radiation and the $\gamma$-rays are described by inverse Compton emission.   After the Fermi-LAT era, the best way to fit the SED of Mrk\,421 was using  a broken power law (two power-law functions).  During the Fermi-LAT multiwavelength campaign  in 2008 - 2010,  it was found that a simple broken power law did not fit adequately the broadband SED  \citep{2011ApJ...736..131A}.  In this work,  three power-law functions (i.e., two breaks) are used to describe the electron population. The double-break power laws can be written as

{\small
\bary\label{espsynm}
\frac{dn_e}{d\gamma_e}&=& N_{0,e}\cases{
\gamma_e^{-\alpha_1}&$\gamma_{\rm min}<\gamma_e\leq \gamma_{\rm c1}$\cr
\gamma_e^{-\alpha_2}   \gamma_{\rm c1}^{\alpha_2-\alpha_1}&$\gamma_{\rm c1}<\gamma_e\leq \gamma_{\rm c2}$\cr
\gamma_e^{-\alpha_3}   \gamma_{\rm c1}^{\alpha_2-\alpha_1}\gamma_{\rm c2}^{\alpha_3-\alpha_2}&$\gamma_{\rm c2}<\gamma_e\leq \gamma_{\rm max}$\,,
}
\eary
}

where $N_{0,e}$ is the number density of electrons, $\alpha_1$,  $\alpha_2$ and  $\alpha_3$ are the spectral indeces and $\gamma_{\rm min}$, $\gamma_{\rm c1}$, $\gamma_{\rm c2}$ and $\gamma_{\rm max}$ are the electron Lorentz factors for minimum, break (1 and 2) and maximum, respectively. It is worth noting that the two broken power laws have three more parameters than a single power law.  In order to describe the  SED,  the one-zone SSC model used in this work follows the methodology presented by  \citet{2016ApJ...830...81F} with some modifications related  to the double-break power laws.  The observed synchrotron spectrum \citep[with an additional power law to that shown in][]{2016ApJ...830...81F} is obtained through the eq. (\ref{espsynm}) and emissivity  $\epsilon_\gamma N_\gamma(\epsilon_\gamma) d\epsilon_\gamma=(- dE_e/dt)\,N_e(E_e)dE_e$ \citep{1994hea2.book.....L, 1986rpa..book.....R}.  For the VHE $\gamma$-ray fluxes, the effect of the extragalactic background light (EBL) absorption modeled by  \cite{2008A&A...487..837F} is introduced.   We did the Chi-square $\chi^2$ minimization using  the ROOT software package \citep{1997NIMPA.389...81B} to fit data and get the values of  the Doppler factor, the size of emitting region, the electron number density and the strength of the magnetic field.  Additionally, the sbezier function was used to smooth the SEDs\footnote{gnuplot.sourceforge.net.}.  With the previous values obtained from the best fit, we estimate several physical properties of  Mrk\,421 in distinct states: the total jet power $L_{\rm jet}$ and energy densities carried by electrons $U_e$, magnetic field $U_B$ and protons $U_p$.  The total jet power is defined by \citep{2008MNRAS.385..283C} $$L_{\rm jet}=\sum_{i=e, p,  B} L_i$$ where {\small $L_i\simeq \pi r_d^2 \Gamma^2 U_i$}  with {\small $U_e=m_e N_e\langle \gamma_e\rangle = m_e\int^{\gamma_{\rm max}}_{\gamma_{\rm min}} \frac{dn_e}{d\gamma_e} d\gamma_e$}, {\small $U_p=N_p m_p$} and {\small $U_B=\frac{B^2}{8\pi}$}.  Here $m_p$ is the proton mass and $\Gamma\approx\delta_D$.

\begin{figure*}
\centering
\includegraphics[width=1.08\textwidth]{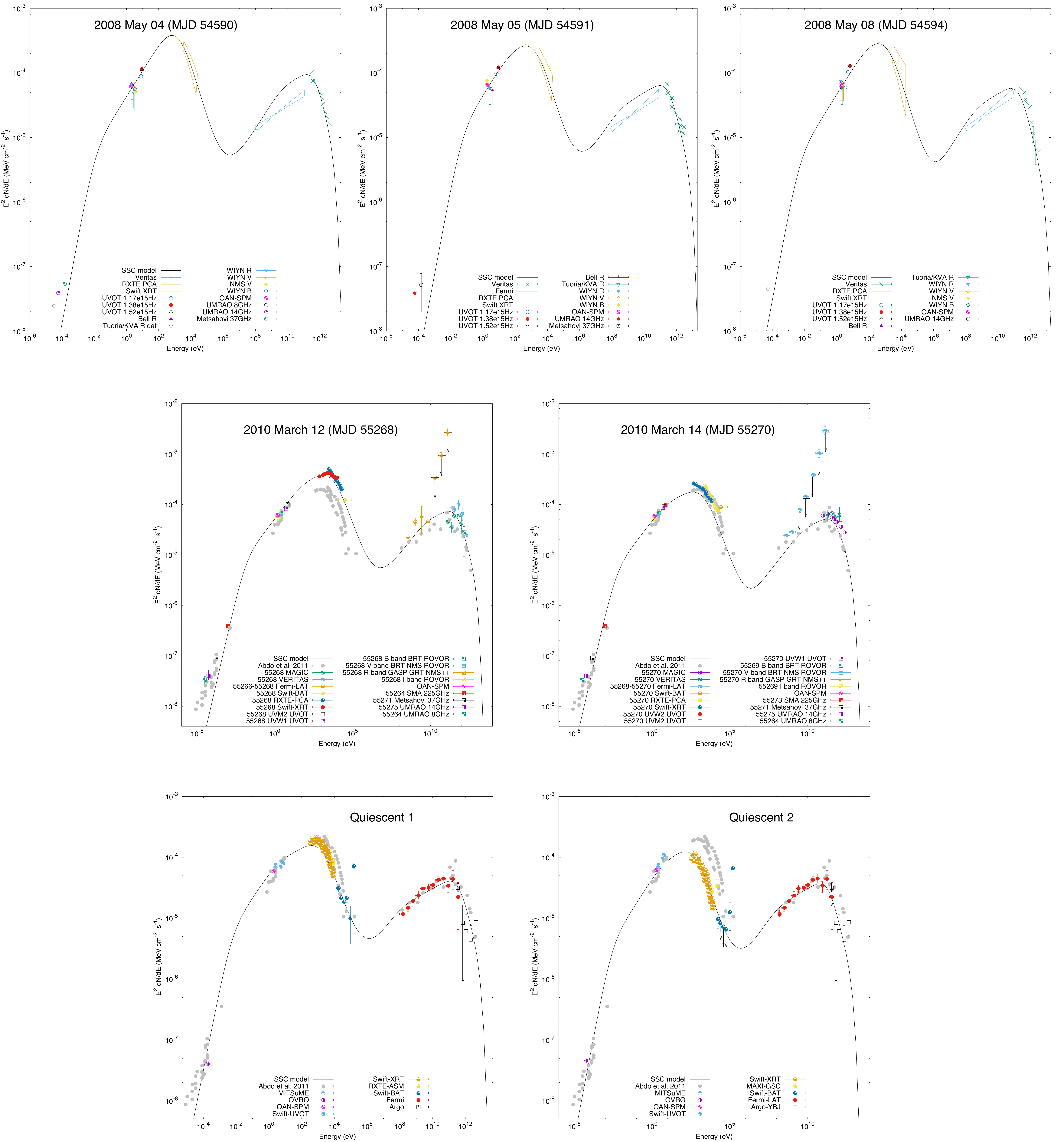}
\caption{The one-zone SSC model was used to fit the SEDs of Mrk\,421 during the flares observed in May 2008, March 2010, and the quiescent states from 2008 August 05 (MJD 54683) to  2009 June 18 (MJD 55000) and from 2010 November 16 (MJD 55516) to 2012 June 28 (MJD 56106).  The best-fit parameters are reported in Table \ref{sed_parameters}}
\label{sed}
\end{figure*}

Figure \ref{sed} shows the fit of the observed SED of Mrk\,421 during the flaring events in May 2008 (top panels) and March 2010 (medium panels), and the quiescent states (bottom panels) from 2008 August 05 (MJD 54683) to 2009 June 18  (MJD 55000) and from 2010 November 16 (MJD 55516) to 2012 June 28 (MJD 56106), respectively.  This figure shows that the one-zone SSC model generated by electron population described through the double-break power laws is successful in describing the SED  during the flaring and quiescent states. The homogeneous one-zone SSC model was used in each panel and the best-fit parameters are reported in Table \ref{sed_parameters}.  This table displays all the parameter values obtained and derived in and from the fit for the values of spectral indexes  $\alpha_1=2.2$,  $\alpha_2=2.7$ and $\alpha_3=4.7$  and $\gamma_{\rm e,min}$ = 800 \citep{2011ApJ...736..131A}.    The values reported in Table \ref{sed_parameters} after fitting are in the range of those obtained by  \cite{2011ApJ...736..131A} and \cite{2015A&A...578A..22A}.     From the values of Doppler factors and emitting radii reported in Table \ref{sed_parameters}, the variability timescales  (eq. \ref{rd}) are $\sim$9.6 and  18.2 hours for the flaring events of May 2008 and March 2010, respectively, and  $\sim$24 hours  for the quiescent states.   The large values of $\gamma_{\rm e,min}$ used in this work to describe the SEDs imply that electrons are efficiently accelerated by the Fermi mechanism only above this energy, and below this energy they are accelerated by a different mechanism that produces the hard electron distribution.  The proton luminosity is computed using the charge neutrality condition to justify a comparable number of electrons and protons \citep[N$_e$=N$_p$;][]{2013ApJ...768...54B, 2009ApJ...704...38S, 2011ApJ...736..131A, 2014A&A...562A..12P,2017APh....89...14F}.  The value of proton luminosity found corresponds to a small fraction ($\sim 10^{-2}$) of the Eddington luminosity $L_{Edd}\sim 2.5\times 10^{46}$ erg/s which is estimated by considering  the value of supermassive black hole ($2\times 10^8 M_\odot$; \citealp{2003ApJ...583..134B}).    The cooling Lorentz factor for radiative electrons {\small $\gamma_c=6\pi m_e/(\sigma_T(1+Y)\, B^2\, r_d)$}  is obtained   by equaling the dynamical and synchrotron cooling times.  The maximum electron Lorentz factor $\gamma_{max}=(3 q_e/\sigma_T)^{1/2}B^{-1/2}$ is calculated by equaling the acceleration and cooling timescales.   Here,  $Y$ is the Compton parameter, as defined in \citet{2016ApJ...830...81F} and $r_{d}$ is defined in eq. \ref{rd}.  Comparing the values of $\gamma_c$ with $\gamma_{c1}$ and $\gamma_{c2}$ shows that the second break in the electron population described by double-break power law is similar to the second break Lorentz factor ($\gamma_{c1}$), probably due to synchrotron cooling. As suggested by \cite{2011ApJ...736..131A} the first break $\gamma_{c1}$  is associated with the acceleration mechanism, and then, energetic electrons above this break are accelerated less efficiently.    Taking into account the values of magnetic field, the electron and the proton luminosities and their ratios $\lambda_{ij}=\frac{L_i}{L_j}$, a principle of equipartition could be present in the jet of Mrk\,421 in order to  relate these luminosities.  The emitting radius found is of the order of $\sim 10^{16}$ cm which is three orders of magnitude larger than the gravitational radius $r_g=5.9\times 10^{13}$ cm.  It is worth noting that the electron energy density is one order of magnitude larger than the magnetic energy density, as reported previously by \cite{2011ApJ...736..131A,2015A&A...578A..22A,2012A&A...542A.100A}.

\subsection{Optical - radio comparison}

Figure \ref{flare_2013} shows the optical flares of April 2013, followed by the increase of the polarization degree and radio flux in May 2013. The polarization degree increased from $\sim$2\% (in April, first period) to  its largest value observed in the R-band of 11\% (in May 2013, during the flare) and brightened also at radio wavelengths. A similar variability behaviour was observed by \cite{1998A&A...339...41T} during the monitoring study done from October 1994 to June 1997. A  large optical flare occurred  from 1996 November 13 to 1997 March 15 and was followed by a radio flare with a delay of 30 - 60 days. In the same observational period the polarization level  increased up to the largest value observed $\sim$12\% in the V-band. \\

\begin{figure*}
\centering
\includegraphics[width=0.9\textwidth]{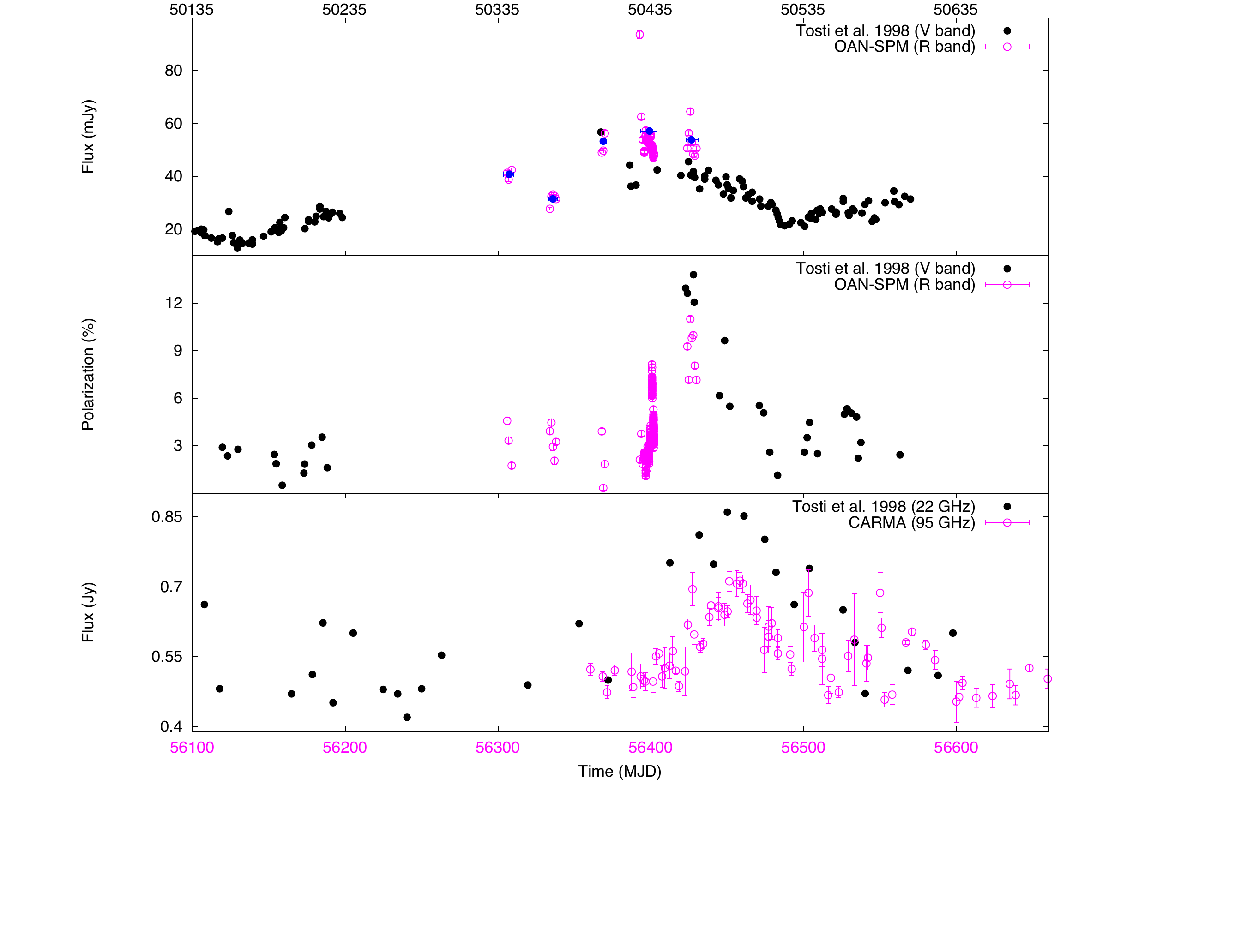}
\caption{Top and bottom panels shows optical (V and R) and radio (22 and 95 GHz) lightcurves, respectively. Middle panel shows the polarization degree variations around the flare observed from 1996 November 13 to 1997 March 15 (black points) and then during April 2013 (magenta points) flare.  The R-band average flux values per run in April 2013 were obtained and appear as blue points in this figure.}
\label{comp_flares}
\end{figure*}

In Figure \ref{comp_flares} the superposition of the optical (V and R bands) and radio (22 and 95 GHz) flares observed in 1996-97 (black points) and April 2013 (magenta points) are shown. This figure clearly shows that the polarization degree (from both epochs) increased up to one of the highest values ever observed.

In order to compare with data obtained by \cite{1998A&A...339...41T}, R-band average flux values per run were obtained and appear as blue points in the figure. The variability behavior of radio flux (bottom panel) and the polarization degree (intermedium panel) present a similar structure.   For instance,  the radio fluxes at 95 GHz and 22 GHz  (bottom panel) present first variations around average values of 0.512$\pm$ 0.003 and  0.514$\pm$ 0.017, respectively, and second,  a rise to a peak, followed by a decay.  The slopes of rise and decay are $(4.26\pm 0.01)\times10^{-3}$ and $(-3.62\pm 0.01)\times10^{-3}$ for the radio flux at 95 GHz and  $(2.54\pm 0.68)\times10^{-3}$  and  $(-3.19\pm 0.65)\times10^{-3}$  for the radio flux at 22 GHz.   Although the optical R-band flux (top panel) is much higher than the V-band ones, they varied in a similar pattern.  In this figure it is observed that both flares present a temporal difference of  $\simeq$ 16.34 years.\\

\subsection{Periodicity analysis of long-term lightcurves}

\begin{figure*}
\centering
\includegraphics[width=0.8\textwidth]{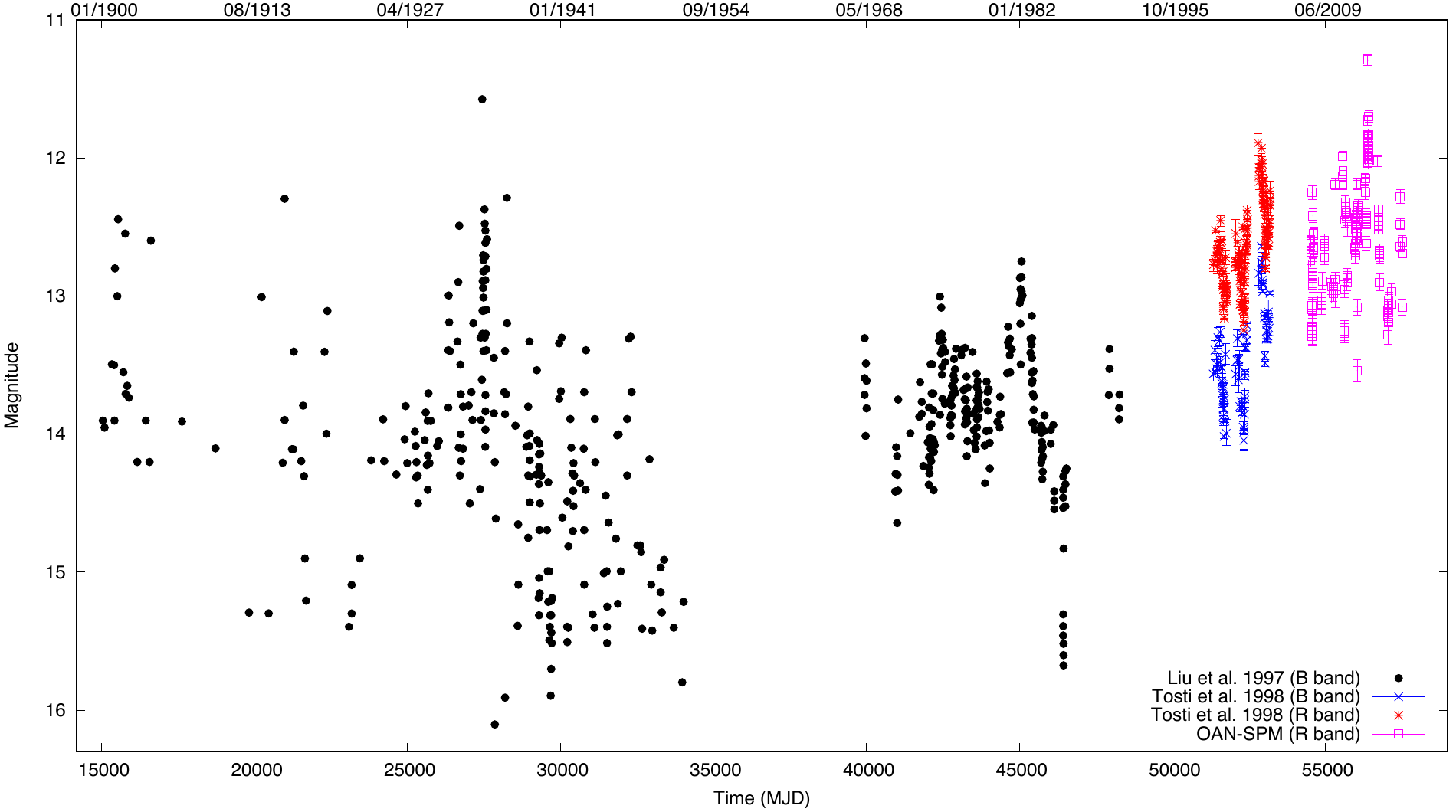}
\caption{The historical light curve of Mrk\,421 from 1900 to 2016.}
\label{hist_opt}
\end{figure*}

In this section the historical light curve of Mrk\,421 obtained from 1900 January 29 to 1991 January 10, which was built by \cite{1997A&AS..123..569L}, will be used along with our optical data in order to look for periodical variations (see Figure \ref{hist_opt}).   Historical data were obtained in the optical B band.  It is worth noting that optical fluxes from different bands (in particular, R and B) are found to be clearly correlated \citep[e.g.][]{1998A&A...339...41T, 2009ApJ...695..596H, 2011ApJ...738...25A, 2015A&A...578A..22A}. 

Periodograms  of the long-term Mrk\,421 lightcurves were obtained by using two different methods: Lomb-Scargle  \citep{1982ApJ...263..835S} and RobPer \citep{Thieler16,2015arXiv151201219B}. 

\begin{figure*}
\centering
\includegraphics[width=\textwidth]{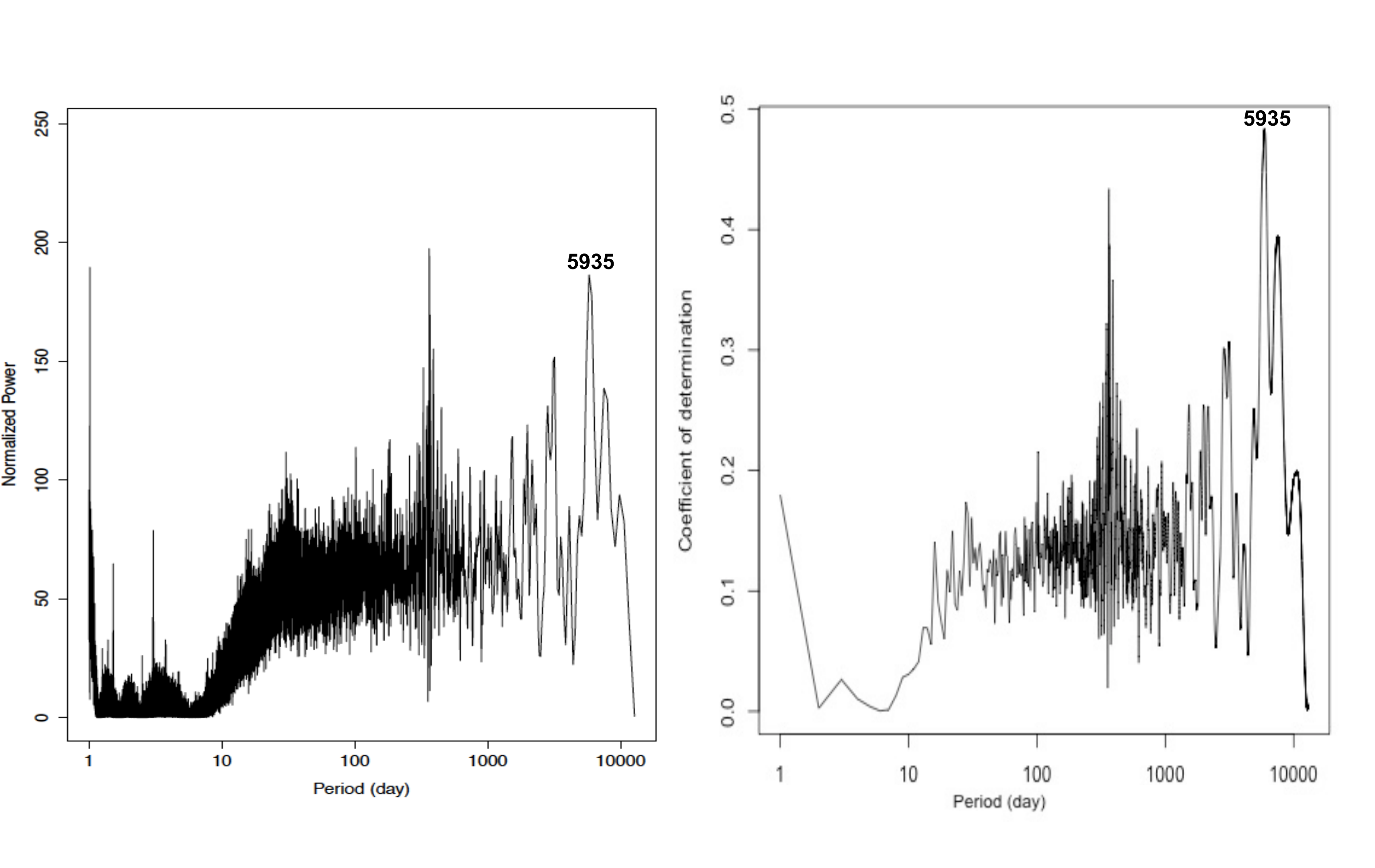}
\caption{Periodograms obtained with the historical light curve data of Mrk\,421 (see fig. \ref{hist_opt}) obtained using Lomb-Scargle (left panel) and RobPer (right panel).}
\label{periods}
\end{figure*}

Using the historical optical data, an analysis with and without  new data were done. The periods found  without the new data are 23.8$\pm$2.1 and 15.9$\pm$1.3  years. These periods are similar to those found by \cite{1997A&AS..123..569L}  which are 23.1$\pm$1.1 and 15.3$\pm$0.7 years, although for these authors the $\sim$23 yr period is more statistically significant.   Adding our data, the $\sim$16 year period result more statistically significant. The periods found with the new data are displayed in Figure \ref{periods}. This Figure shows the analysis done with the Lomb-Scargle (left panel) periodogram and also with the R package RobPer (right panel). The analysis with  Lomb-Scargle and RobPer methods provides the same periodicities of 1, 365 and 5935 days.The periods of one day and 365 days are due to the window function \citep{2010ApJ...722..937D}.  Therefore, only the period of 5935 $\pm$650 days is real and it is in agreement with one of the periods found by \cite{1997A&AS..123..569L}. This period naturally explains the similar flares observed in radio and optical bands of 1996-97 and 2013, i.e. $\simeq$16 years later.  
%On the other hand, the period found with the new data is 16.3 $\pm$ 1.8 years.

\subsection{V and R spectral index variability analysis}

\begin{figure*}
\centering
\includegraphics[width=0.9\textwidth]{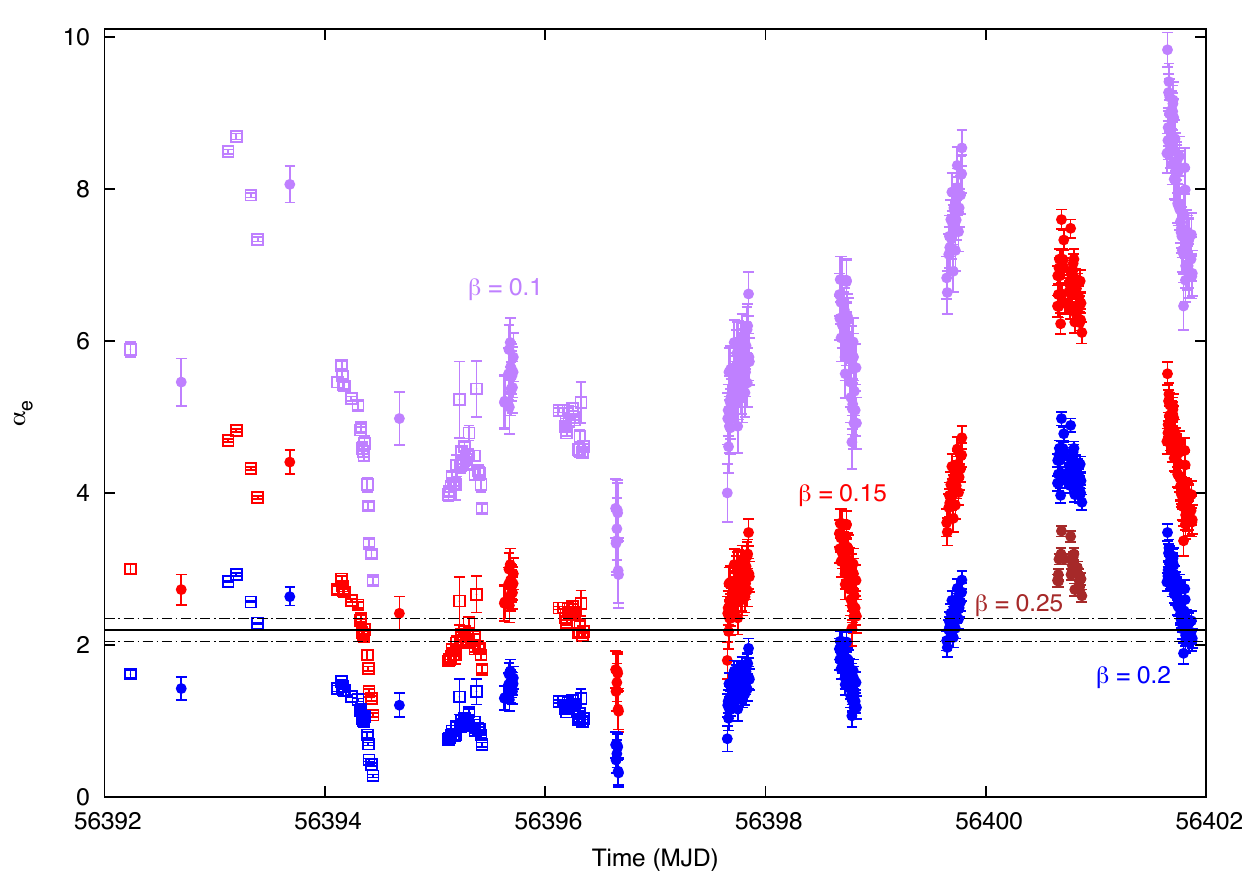}
\caption{The intranight spectral index variations as a function of time for different values of $\beta$. The  V \citep[$\Box$;][]{2016A&A...591A..83S}  and R ($\bullet$) optical bands.}
\label{alpha_color}
\end{figure*}

Figure \ref{alpha_color} shows the intranight spectral index variations as a function of time for different values of $\beta$. In this figure  the  V and R optical bands are considered when Mrk\,421 exhibited the unprecedented flaring activity from April 09 (MJD 56391) to 19.  The V-band optical data used in this section are reported in \cite{2016A&A...591A..83S}.   Electrons in a magnetized plasma  are usually cooled down by synchrotron radiation, then cooling timescales of electrons in a few hours could be related with variations in the V and R optical bands. In fact, the changes in the V and R optical bands together with the observed  intranight flux variability suggest that the acceleration timescale is less than the cooling timescale in the emitting region, and the particle acceleration occurs at the shock front, with a magnetic field produced by the plasma compression \citep{1985ApJ...298..114M}.   Considering a variability timescale of five hours  ($\tau_{var}\sim$ 5 hours, see Table \ref{daily_var}), and a typical value of Doppler factor  $\delta_D\sim$ 20 (see Table 9), then the magnetic field and the emitting region become  $B\lesssim$ 1.2 G and $r_d=9.8\times 10^{15}$ cm, respectively.  These values are in agreement with those reported by \cite{2011ApJ...736..131A}.

Considering the scattering of the polarization degree $\delta P$ and the emitting region $r_d$, then the coherence length of the large-scale field can be estimated as \citep{1985ApJ...290..627J}

\be\label{lB}
l_B=\left(\frac{k\,\Pi_0\,\delta P}{r_d^{3/2}}\right)^{-2/3}\,.
\ee

Taking into account the values of  $\delta P\sim$2 \%  (see Fig.  \ref{alpha_color} for $\alpha=2.2$),  k=0.5 \citep{2013ApJS..206...11S}, $\Pi_0$=0.705,  $r_d\simeq 9.8\times 10^{15}$ cm,  then the coherence length is $l_B$=0.26 pc.   Taking into account  the intranight variation  associated with the distance travelled by the relativistic shocks, we can compare the  linear scale  (eq \ref{lB}) with this distance.   Following  \citet{1967MNRAS.135..345R} and \cite{1991A&A...241...15Q},  we estimate the distance travelled by the relativistic shocks using

\be
D= \frac{\beta_s\,\delta_s\,\Gamma_s\,\Delta t} {(1+z)}\,,
\ee

where $\beta_s$ is the shock speed,  $\Delta$t is the observed timescale \citep{1991A&A...241...15Q}  and $\Gamma_s$ is the Lorentz factor.  Using the values of  shock speed given in \cite{2004ApJ...600..115P} and \cite{2008ApJ...678...64P}, the distance travelled by the shocks is 0.29 pc.  The good agreement found between the values of the distance travelled by shocks along the jet and the field turbulence scale encourage us to think  that intranight variations  observed in the polarization degree and spatial changes in the magnetic field could be strongly correlated with  inhomogeneities.  These  inhomogeneities could have their origin in  the jet or in the plasma by compressing and re-ordering the complex structure of the magnetic field \citep{1980MNRAS.193..439L}.

\section{Conclusions}

Long-term R-band photopolarimetric observations of the HBL  Mrk\,421 carried out since February 2008 up to May 2016 have been presented.   The highest brightness state of 11.29$\pm$0.03 mag (93.60$\pm$1.53 mJy) and the polarization degree value of 11.00$\pm$0.44\% have never been detected before in the R-band lightcurves.  Extensive multiwavelength observations covering radio to TeV $\gamma$-rays around the flares observed in May 2008, March 2010,  and April 2013 have been studied.   From the photopolarimetric analysis presented in this work in the  optical R-band the following results are found.

From the flaring activity in May 2008, we have shown that the optical flux varies on timescales of 15.4 days and that it is anti-correlated with TeV $\gamma$-rays, hard/soft X-rays, the polarization degree and the position angle. In addition, a strong correlation between the polarization degree and the position angle was found.  These results suggest that Mrk\,421 could have different emitting regions which might play a role in different situations. In March 2010, a correlation  among  TeV $\gamma$-ray, X-ray and optical fluxes was found during the high activity state, in agreement with the one-zone SSC model.

The most important results were found in 2013 where this object displayed the highest activity levels.  In order to do a more detail analysis, the flare observed in  April 2013  was divided in three periods.   During the first period, Mrk\,421 presented very high activity in TeV, soft X-ray and optical bands.  The optical and TeV $\gamma$-ray  fluxes  were found to be anti-correlated, and the polarization degree and the position angle were found to be correlated.  During the second period, a large rotation of the position angle of $\sim\, 100^\circ$ was detected when the degree of polarization was $\sim 2\%$. This result could be due to shocks traveling along helical magnetic field lines  \citep{2008Natur.452..966M}.  The TeV $\gamma$-ray  and hard/soft X-ray flares were accompanied by a modest increase in optical activity.  The optical flux was found to be strongly correlated with the position angle and anti-correlated with the polarization degree.  Additionally, a strong anti-correlation between the polarization degree and the position angle was found.  In the third period, a maximum value of the polarization degree  of $\sim\, 8.33\,\%$ was detected when TeV $\gamma$-ray, X-ray and optical fluxes were decreasing.  No strong correlations were found among the optical flux with the rest of the higher energy fluxes. The normalized Stokes parameters $q$ and $u$ were found strongly correlated,  thus suggesting that the observed variability  was due to a single variable component with constant polarization properties \citep{2008ApJ...672...40H}.

The entire data set obtained in 2013 were used to analyse the Stokes parameters. The superposition of two polarized components coming from distinct emitting regions were found: one constant polarized component and other variable one. The constant component had a polarization degree of $P=3.25\pm 0.70$\% and a position angle of  $\theta_{\rm con}=219.01^\circ\pm0.02$.  The component that accounts for the variability behaviour can be naturally explained due to the propagation of the shocks \citep{1984MNRAS.211..497H, 1996MNRAS.282..788B}.

The first intranight photopolarimetric variability study in the R-band of Mrk\,421 from 2013 April 13 to 19 is presented in this work. For the OAN-SPM data the following results were found:

\begin{enumerate}
\item The minimum variability timescale of the flux is found to be  $\sim$ 2.3 hours whereas the polarization degree and the spectral index vary on timescales of $\Delta t\sim$ minutes.   Variations of the polarization degree could be related to the propagation of the shocks by means of an inhomogeneous plasma, compressing and re-ordering the magnetic field lines \citep{1980MNRAS.193..439L}.
\item From night to night, the polarization degree variability presented variations  of $\sim$ 2 - 3 $\%$. In addition, for the spectral index in the range of 2 $\leq\alpha_e\leq$ 3 \cite[see,][]{2011ApJ...736..131A}, the ratio of ordered to chaotic magnetic field intensity was found to be in the range of 0.1 $\leq\beta\leq$ 0.3.  These variations might be explained when the shocks travel through  the jet and find different regions with distinct magnetic field properties, leading a changing ratio of ordered to chaotic magnetic field intensities \citep{2010MNRAS.408.1778B}.
\item The correlation found between the normalized Stokes parameters (q and u)  analyzed per night showed a temporal evolution.  It could suggest that the total polarized flux is produced by more than one polarized component \citep{2008ApJ...672...40H}.
\end{enumerate}

Using the 100-year-historical light curve of Mrk\,421 and our SPM data,  a periodicity of 5935$\pm$650 days (16.26 $\pm$1.78 years) was found.  One possible explanation for the  periodicity is given through the dynamics of an orbiting binary supermassive black hole system and their accretion disks.

In addition, the analysis of the radio and photopolarimetric lightcurves observed during the flares of 1996-97 and  $\simeq$ 16.34 years later in 2013  indicate that both flares could have been produced in the same emitting region.  This result has important implications for theoretical models that intend to explain the optical (with polarimetry) and radio variability of this source. These results will be analysed in a forthcoming paper.

The broadband SEDs were fitted with the one-zone SSC model. An electron population with three power-law functions was required to describe the SEDs during the flaring and quiescent states.  The full set of parameters derived by using the SED are: Doppler factor $\delta_D=$ 20 - 24, magnetic field $B=  (34.3 -  43.1)  \times 10^{-3}$ G, emitting radius $r_d=$ (2.5 - 5.0) $\times 10^{16}$ cm and electron density $N_e=$(0.1 -  0.26) cm$^{-3}$. These values are in agreement with those reported by  \cite{2011ApJ...736..131A,2015A&A...578A..22A,2012A&A...542A.100A}.  The values  of the magnetic field and the electron density found are higher in flaring than in quiescent states.  This indicates that the radiative efficiency of electrons is higher in flaring events.  The fit applied to the broadband SEDs yields the variability timescales that range from 9 hours to one day.    The largest value of one day corresponds to the description of the quiescent states and the smaller value (9 hours) is related with the flaring event.  This shows that the variability timescale could be related with the state of activity and then, with the acceleration and the cooling mechanisms produced by the electron population.

A detailed analysis of the electron spectral index with the  V and R optical bands was performed. The results indicate that the coherence length of the large-scale magnetic field, and the distance travelled by the relativistic shocks are of the same order, thus suggesting a connection between the intranight variations observed in the polarization degree and the spatial changes in the magnetic field.

\section*{Acknowledgements}
We thank the anonymous referee for a critical reading of the
paper and valuable suggestions that helped improve the quality
and  clarity  of  this  work.  E.B., M.R. and F.S.  acknowledge  financial  support  from UNAM-DGAPA-PAPIIT  through  grant  IN111514  and N.F. through grant  IA102917.  D.H. also acknowledge support from CONACyT through grant 180817.   N.F.,  E.B. and D.H.  thank support granted by CONACyT: 233107, 13654 and 8366.   D.H. thanks PASPA-DGAPA UNAM for the fellowship to spend a sabbatical year at the LSW, Heidelberg University. This work is based upon observations carried out at the Observatorio Astron\'omico Nacional on the Sierra San Pedro M\'artir (OAN-SPM), Baja California, Mexico. We also acknowledge Patrick Moriarty, Matthias Beilicke, Chen Songzhan, Atreyee Sinha, Deirdre Horan,  Maria Petropoulou and  Talvikki  Hovatta for sharing with us part of the data used in this work. This research has made use of the SAO/NASAAstrophysics Data System (ADS) and of the NASA/IPAC Extragalactic Database (NED), which is operated by the Jet Propulsion Laboratory, California Institute of Technology, under contract with the National Aeronautics and Space Administration.  We acknowledge the use of the Fermi-LAT publicly available data as well as the public data reduction software. 

%%%%%%%%%%%%%%%%%%%%BIBLIOGRAPHY%%%%%%%%%%%%%%%%%%%%%%%%%%%%%%%%%%%%%%
\clearpage
%\bibliography{Bib_mrk421}
%\addcontentsline{toc}{chapter}{Bibliography}

\clearpage
%%%%%%%%%%%%%%%%%%%%%%%%%%%%%%%%%%%%%%%%%%%%%%%%%%%%%%%%%%%%%%%%%%%%%%%%

\begin{table*}
\begin{center}\renewcommand{\arraystretch}{1.5}\addtolength{\tabcolsep}{4pt}
\caption{R-band photopolarimetric data of Mrk\,421 from February 2008 to May 2016}\label{all_data}
\begin{tabular}{ c c c c c c c c c}
\hline \hline
 MJD           &     P   &   $\Delta$P  &     PA          & $\Delta$PA  & R$_{\rm mag}$         &  $\Delta$R$_{\rm mag}$ &  $F_{\rm R}$& $\Delta F_{\rm R}$\\
   &  (\%) &   (\%)           &  ($^\circ$)   &   ($^\circ$)   &   &         & (mJy)& (mJy)\\
  (1)            &   (2)    &      (3)          &      (4)         &      (5)           &   (6)    &    (7)            &  (8)   & (9)\\          
 \hline\hline
54524.9126&   1.51&  0.17&   177.1&   2.8&   12.61&   0.05&   27.728&  1.265\\
54524.9185&   1.84&  0.17&   181.4&   2.3&   12.62&   0.05&   27.548&  1.260\\
54525.8853&   3.01&  0.18&   166.2&   1.5&   12.74&   0.05&   24.659&  1.175\\

54525.8901&   3.03&  0.18&   166.3&   1.5&   12.74&   0.05&   24.686&  1.175\\
54567.8052&   1.62&  0.20&   172.7&   2.9&   13.10&   0.06&   17.776&  0.989\\
54567.8105&   1.78&  0.20&   152.6&   2.2&   13.10&   0.06&   17.672& 0.986\\
54569.6460&   3.21&  0.20&   211.8&   1.4&   13.08&   0.06&   18.093&  0.998\\
....& ....& ....& ....& ....& ....& ....& ....& ....\\
\hline
 \end{tabular}
\end{center}
\begin{center}
 A portion of the whole data is shown here for guidance regarding its form and content. The complete data of Table 1 are available in a machine-readable form in the on-line journal. Cols. (2) and (3): values and associated errors of polarization degree, cols. (4) and (5) values and associated errors of position angle, cols. (6) and (7) values and associated errors of magnitude, cols. (8) and (9) values and associated errors of optical flux.\\
\end{center}
\end{table*}
%%%%%%%%%%%%%%%%%%%%%%%%%%%%%%%%%%%%%%%%%%%%%%%%%%%%%%%%%%%%%%%%%%%%%%%%%
%
\begin{table*}
\begin{center}\renewcommand{\arraystretch}{2}\addtolength{\tabcolsep}{6pt}
\caption{The maximum and minimum values of the R-band photopolarimetric observations}\label{gen_data}
\vspace{0.1cm}
\begin{tabular}{ c c c c c}
\hline
\hline
 \normalsize{Parameter}& \normalsize{Max/Date}&\normalsize{Min/Date} & \normalsize{Average} & \normalsize{$\chi^2$} \\
 (1)            &   (2)    &      (3)          &      (4)         &      (5) \\ 
\hline
\hline
 \normalsize{$F_{\rm R}$(mJy)}                          &  \scriptsize{($93.60\pm1.53$) / (10-04-2013) }        &     \scriptsize{($11.81\pm0.38$) / (24-04-2012)}         &       \scriptsize{$35.09\pm0.74$} &  \scriptsize{$1.72\times 10^5$} \\
\normalsize{$R_{mag}$}                          &  \scriptsize{($11.29\pm0.04$) / (10-04-2013)}        &     \scriptsize{($13.54\pm0.08$) / (24-04-2012)}         &       \scriptsize{$12.10\pm0.02$} & \scriptsize{$2.31\times 10^4$}  \\
\normalsize{P(\%)}                         &  \scriptsize{($11.00\pm0.18$) / (13-05-2013)}        &     \scriptsize{($0.35\pm0.17$) / (17-03-2013)}         &       \scriptsize{$3.62\pm0.10$}  & \scriptsize{$3.32\times 10^4$}   \\
\normalsize{$\theta(\degr)$}              &  \scriptsize{($285.4\pm8.1$) / (27-03-2014)}        &     \scriptsize{($10.0\pm4.2$) / (24-02-2012)}         &       \scriptsize{$168.83\pm2.19$} &  \scriptsize{$3.33\times 10^5$}  \\
\hline
\hline
\end{tabular}
\end{center}
\begin{center}
Cols (2) and (3) show the maximum and minimum values (observational dates in parenthesis) of the optical flux, R-band magnitude,  polarization degree and position angle.  Columns (4) shows the average values  during the 8 years of observations and Column (5) shows the Chi square values. \\
\end{center}
\end{table*}
%
%%%%%%%%%%%%%%%%%%%%%%%%%%%%%%%%%%%%%%%%%%%%%%%%%%%%%%%%%%%%%%%%%%%%%%%%%
%
\begin{table*}
\begin{center}\renewcommand{\arraystretch}{1.6}\addtolength{\tabcolsep}{6pt}
\caption{Variability of the statistical parameters}\label{stat_year}
\begin{tabular}{ c c c c c c c c}
\hline
\hline
\normalsize{Year} & \normalsize{Parameter}& \normalsize{Max}&\normalsize{Min} & \normalsize{Average}& \normalsize{Y(\%)}&\normalsize{$\mu(\%)$}& \normalsize{$\cal F $} \\
(1)            &   (2)    &      (3)          &      (4)         &      (5)           &   (6)    &    (7)            &  (8) \\
\hline
\hline
			    & \normalsize{$F_{\rm R}$(mJy)}                          &  \scriptsize{38.94$\pm$1.62}        &     \scriptsize{14.86$\pm$0.92}         &       \scriptsize{$23.39\pm0.09$}      &    \scriptsize{102.80}          &  \scriptsize{0.40}  &  \scriptsize{0.45}                \\
\normalsize{2008} & \normalsize{P(\%)}                         &  \scriptsize{7.97$\pm$0.28}        &     \scriptsize{1.28$\pm$0.29}         &       \scriptsize{$3.29\pm0.05$}      &    \scriptsize{102.41}          &  \scriptsize{1.39}  &  \scriptsize{0.72}              \\
			    &\normalsize{$\theta(\degr)$}              &  \scriptsize{211.8$\pm$1.4}        &     \scriptsize{102.2$\pm$1.2}         &       \scriptsize{$176.67\pm0.45$}      &    \scriptsize{62.02}          &  \scriptsize{0.26}    &  \scriptsize{0.35}               \\
\hline
			    & \normalsize{$F_{\rm R}$(mJy)}                          &  \scriptsize{28.21$\pm$0.56}        &     \scriptsize{18.26$\pm$0.44}         &       \scriptsize{$24.58\pm0.17$}      &    \scriptsize{40.28}          &  \scriptsize{0.70}  &  \scriptsize{0.21}                \\
\normalsize{2009} & \normalsize{P(\%)}                         &  \scriptsize{7.43$\pm$0.28}        &     \scriptsize{2.01$\pm$0.27}         &       \scriptsize{$4.06\pm0.11$}      &    \scriptsize{132.89}          &  \scriptsize{2.57}  &  \scriptsize{0.57}               \\
			    &\normalsize{$\theta(\degr)$}              &  \scriptsize{156.3$\pm$1.5}        &     \scriptsize{118.1$\pm$1.4}         &       \scriptsize{$137.33\pm0.65$}      &    \scriptsize{27.74}          &  \scriptsize{0.48}  &  \scriptsize{0.14}                 \\
\hline
			    & \normalsize{$F_{\rm R}$(mJy)}                          &  \scriptsize{41.01$\pm$0.73}        &     \scriptsize{19.16$\pm$0.45}         &       \scriptsize{$22.74\pm0.17$}      &    \scriptsize{95.92}          &  \scriptsize{0.73}   &  \scriptsize{0.36}               \\
\normalsize{2010} & \normalsize{P(\%)}                         &  \scriptsize{7.53$\pm$0.30}        &     \scriptsize{1.25$\pm$0.29}         &       \scriptsize{$3.77\pm0.09$}      &    \scriptsize{166.09}          &  \scriptsize{2.51}     &  \scriptsize{0.72}           \\
			    &\normalsize{$\theta(\degr)$}              &  \scriptsize{180.3$\pm$2.8}        &     \scriptsize{114.3$\pm$4.0}         &       \scriptsize{$131.84\pm0.83$}      &    \scriptsize{49.78}          &  \scriptsize{0.63}  &  \scriptsize{0.22}                \\
\hline
			    & \normalsize{$F_{\rm R}$(mJy)}                          &  \scriptsize{49.05$\pm$0.85}        &     \scriptsize{15.17$\pm$0.40}         &       \scriptsize{$31.68\pm0.16$}      &    \scriptsize{106.83}          &  \scriptsize{0.49}  &  \scriptsize{0.53}               \\
\normalsize{2011} & \normalsize{P(\%)}                         &  \scriptsize{5.53$\pm$0.43}        &     \scriptsize{1.49$\pm$0.26}         &       \scriptsize{$3.75\pm0.07$}      &    \scriptsize{106.11}          &  \scriptsize{1.93}    &  \scriptsize{0.58}              \\
			    &\normalsize{$\theta(\degr)$}              &  \scriptsize{178.6$\pm$2.6}        &     \scriptsize{16.4$\pm$2.4}         &       \scriptsize{$89.17\pm0.49$}      &    \scriptsize{181.36}          &  \scriptsize{0.55}   &  \scriptsize{0.83}               \\
\hline
			    & \normalsize{$F_{\rm R}$(mJy)}                          &  \scriptsize{40.96$\pm$0.73}        &     \scriptsize{11.81$\pm$0.37}         &       \scriptsize{$30.13\pm0.11$}      &    \scriptsize{96.68}          &  \scriptsize{0.37}   &  \scriptsize{0.55}               \\
\normalsize{2012} & \normalsize{P(\%)}                         &  \scriptsize{6.56$\pm$0.21}        &     \scriptsize{0.63$\pm$0.24}         &       \scriptsize{$2.60\pm0.05$}      &    \scriptsize{229.24}          &  \scriptsize{1.78}       &  \scriptsize{0.83}           \\
			    &\normalsize{$\theta(\degr)$}              &  \scriptsize{167.6$\pm$8.1}        &     \scriptsize{10.0$\pm$4.2}         &       \scriptsize{$65.65\pm0.58$}      &    \scriptsize{239.24}          &  \scriptsize{0.88}     &  \scriptsize{0.89}           \\
\hline
			    & \normalsize{$F_{\rm R}$(mJy)}                          &  \scriptsize{93.60$\pm$1.53}        &     \scriptsize{27.70$\pm$0.55}         &       \scriptsize{$51.46\pm0.06$}      &    \scriptsize{127.94}          &  \scriptsize{0.11}      &  \scriptsize{0.54}           \\
\normalsize{2013} & \normalsize{P(\%)}                         &  \scriptsize{11.00$\pm$0.18}        &     \scriptsize{0.35$\pm$0.17}         &       \scriptsize{$3.77\pm0.12$}      &    \scriptsize{282.64}          &  \scriptsize{0.31}          &  \scriptsize{0.94}       \\
			    &\normalsize{$\theta(\degr)$}              &  \scriptsize{217.8$\pm$0.8}        &     \scriptsize{31.0$\pm$1.7}         &       \scriptsize{$177.87\pm0.11$}      &    \scriptsize{105.01}          &  \scriptsize{0.06}   &  \scriptsize{0.76}              \\
\hline
			    & \normalsize{$F_{\rm R}$(mJy)}                          &  \scriptsize{48.01$\pm$0.84}        &     \scriptsize{21.22$\pm$0.47}         &       \scriptsize{$31.11\pm0.15$}      &    \scriptsize{86.01}          &  \scriptsize{0.49}    &  \scriptsize{0.39}              \\
\normalsize{2014} & \normalsize{P(\%)}                         &  \scriptsize{7.03$\pm$0.30}        &     \scriptsize{0.65$\pm$0.27}         &       \scriptsize{$2.83\pm0.08$}      &    \scriptsize{224.83}          &  \scriptsize{2.89}            &  \scriptsize{0.83}     \\
			    &\normalsize{$\theta(\degr)$}              &  \scriptsize{285.4$\pm$8.1}        &     \scriptsize{160.5$\pm$1.3}         &       \scriptsize{$223.76\pm1.31$}      &    \scriptsize{55.58}          &  \scriptsize{0.59}    &  \scriptsize{0.28}             \\ 
\hline
			    & \normalsize{$F_{\rm R}$(mJy)}                          &  \scriptsize{19.97$\pm$0.46}        &     \scriptsize{15.02$\pm$0.40}         &       \scriptsize{$17.70\pm0.12$}      &    \scriptsize{27.52}          &  \scriptsize{0.70}     &  \scriptsize{0.14}             \\
\normalsize{2015} & \normalsize{P(\%)}                         &  \scriptsize{9.42$\pm$0.33}        &     \scriptsize{3.39$\pm$0.29}         &       \scriptsize{$6.52\pm0.09$}      &    \scriptsize{91.93}          &  \scriptsize{1.36}            &  \scriptsize{0.47}     \\
			    &\normalsize{$\theta(\degr)$}              &  \scriptsize{168.90$\pm$1.30}        &     \scriptsize{149.10$\pm$1.80}         &       \scriptsize{$157.92\pm0.37$}      &    \scriptsize{12.38}          &  \scriptsize{0.23}       &  \scriptsize{0.06}          \\ 
\hline
			    & \normalsize{$F_{\rm R}$(mJy)}                          &  \scriptsize{37.82$\pm$0.77}        &     \scriptsize{18.05$\pm$0.40}         &       \scriptsize{$28.01\pm0.21$}      &    \scriptsize{70.44}          &  \scriptsize{0.75}          &  \scriptsize{0.35}        \\
\normalsize{2016} & \normalsize{P(\%)}                         &  \scriptsize{7.91$\pm$0.32}        &     \scriptsize{0.83$\pm$0.25}         &       \scriptsize{$3.43\pm0.11$}      &    \scriptsize{205.74}          &  \scriptsize{3.19}         &  \scriptsize{0.81}        \\
			    &\normalsize{$\theta(\degr)$}              &  \scriptsize{129.6$\pm$1.6}        &     \scriptsize{47.5$\pm$4.6}         &       \scriptsize{$101.11\pm1.23$}      &    \scriptsize{80.91}          &  \scriptsize{1.21}     &  \scriptsize{0.46}            \\ 
\hline
\hline
\end{tabular}
\end{center}
\begin{center}
Column (2) $F_{R}$ is the R-band flux, P(\%) the polarization percentage and $\theta(\deg)$ the position angle. Columns (3), (4) and (5) show the maximum, minimum and average values of the parameters. Columns (6), (7) and (8) are  the amplitude of the variations, the fluctuation index and the fractional variability index.\\
\end{center}
\end{table*}
%%%%%%%%%%%%%%%%%%%%%%%%%%%%%%%%%%%%%%%%%%%%%%%%%%%%%%%%%%%%%%%%%%%%%%%%%
%
\begin{table*}
\begin{center}\renewcommand{\arraystretch}{1.6}\addtolength{\tabcolsep}{6pt}
%\vspace{4cm}
\caption{Pearson's correlation coefficients}\label{pearson_year}
\begin{tabular}{ c c c c c c c c}
\hline
\hline
\normalsize{Year} &  \normalsize{$F_{\rm R}$-P}&\normalsize{$F_{\rm R}$-$\theta$} & \normalsize{P-$\theta$}& \normalsize{$F_{\rm R}$-u}&\normalsize{$F_{\rm R}$-q}& \normalsize{u-q} \\
(1)            &   (2)    &      (3)          &      (4)         &      (5)           &   (6)    &    (7)       \\
\hline
\hline
\normalsize{2008} &\scriptsize{0.37 (0.11)}          &     \scriptsize{-0.19 (0.44)}         &       \scriptsize{-0.51 (0.03)}      &    \scriptsize{0.12 (0.65)}          &  \scriptsize{-0.26 (0.30)}  &  \scriptsize{-0.44 (0.05)}              \\
\normalsize{2009} &  \scriptsize{-0.03 (0.94)}        &     \scriptsize{-0.30 (0.43)}         &       \scriptsize{0.35 (0.36)}      &    \scriptsize{0.01 (0.98)}          &  \scriptsize{-0.29 (0.47)}  &  \scriptsize{-0.40 (0.27)}               \\
\normalsize{2010} &  \scriptsize{-0.04 (0.92)}        &     \scriptsize{-0.18 (0.64)}         &       \scriptsize{-0.24 (0.53)}      &    \scriptsize{-0.41 (0.29)}          &  \scriptsize{-0.41 (0.29)}     &  \scriptsize{0.88 (2.1$\times10^{-3}$)}           \\
\normalsize{2011}  &  \scriptsize{-0.40 (0.13)}        &     \scriptsize{0.04 (0.88)}         &       \scriptsize{-0.52 (0.04)}      &    \scriptsize{-0.38 (0.17)}          &  \scriptsize{-0.45 (0.06)}    &  \scriptsize{0.64 (0.01)}              \\
\normalsize{2012}  &  \scriptsize{0.24 (0.22)}        &     \scriptsize{0.05 (0.80)}         &       \scriptsize{-0.45 (0.02)}      &    \scriptsize{0.44 (0.01)}          &  \scriptsize{0.36 (0.11)}       &  \scriptsize{0.33 (0.06)}           \\
\normalsize{2013}  &  \scriptsize{-0.15 (0.02)}      &     \scriptsize{0.20 (1.2$\times10^{-3}$)}         &       \scriptsize{0.32 (1.4$\times10^{-7}$)}      &    \scriptsize{0.08 (0.19)}          &  \scriptsize{-0.09 (0.15)}          &  \scriptsize{0.64 ($5.6\times 10^{-9}$)}       \\
\normalsize{2014}  &  \scriptsize{0.34 (0.31)}        &     \scriptsize{-0.21 (0.54)}         &       \scriptsize{-0.38 (0.25)}      &    \scriptsize{0.51 (0.12)}          &  \scriptsize{-0.45 (0.18)}            &  \scriptsize{-0.62 (0.05)}     \\
\normalsize{2015}  &  \scriptsize{-0.69 (0.01)}        &     \scriptsize{-0.16 (0.62)}         &       \scriptsize{-0.09 (0.78)}      &    \scriptsize{-0.63 (0.04)}          &  \scriptsize{0.41 (0.17)}            &  \scriptsize{-0.51 (0.08)}     \\
\normalsize{2016}  &  \scriptsize{0.40 (0.38)}        &     \scriptsize{-0.53 (0.22)}         &       \scriptsize{0.49 (0.26)}      &    \scriptsize{0.35 (0.47)}          &  \scriptsize{0.37 (0.39)}         &  \scriptsize{0.71 (0.06)}        \\
\hline
\hline
\end{tabular}
\end{center}
\begin{center}
Col (2) Optical flux and polarization degree, col (3) optical flux and position angle, col (4)  polarization and position angle. Cols (5) and (6) optical flux and normalized Stokes parameters $u$ and $q$,  and col (7) the normalized Stokes parameters $q$ and $u$.  From col (2) to col (7) numbers in parenthesis are the corresponding p values.\\
\end{center}
\end{table*}
%%%%%%%%%%%%%%%%%%%%%%%%%%%%%%%%%%%%%%%%%%%%%%%%%%%%%%%%%%%%%%%%%%%%%%%%%
%
\begin{table*}
\begin{center}\renewcommand{\arraystretch}{1.6}\addtolength{\tabcolsep}{3pt}
\caption{Pearson's correlation coefficients}\label{pearson_all}
\begin{tabular}{ c c c c c c c}
\hline
\hline
\normalsize{Observables} & \normalsize{$r_{08}$} & \normalsize{$r_{10}$} & \normalsize{$r_{\rm 13, ep1}$}& \normalsize{$r_{\rm 13, ep2}$}& \normalsize{$r_{\rm 13, ep3}$} & \normalsize{$r_{\rm 13, ep4}$}\\
(1)            &   (2)    &      (3)          &      (4)         &      (5)           &   (6)    &    (7)       \\
\hline
\normalsize{$F_{\rm R}$ - $F_{\rm TeV}$}  &  \scriptsize{$-0.88$ (0.02)}  &  \scriptsize{$0.98$ (0.02)}  &  \scriptsize{$-0.82$ (0.39)}  &  \scriptsize{$0.09$ (0.69)}  &  \scriptsize{$0.54$ ($1.6\times 10^{-8}$)}  &  \scriptsize{$-$}   \\
\normalsize{$F_{\rm R}$ - $F_{\rm X, s}$}  &  \scriptsize{$-0.98$ (5.9$\times10^{-4}$)}  &  \scriptsize{$0.98$ (0.02)} & \scriptsize{$-0.93$ (0.24)}  &  \scriptsize{$-0.84$  (1.0$\times10^{-6}$)}  &  \scriptsize{$0.48$ ($3.2\times10^{-8}$)}  &  \scriptsize{$1.0$ (1.9$\times10^{-5}$)}  \\
\normalsize{$F_{\rm R}$ - $F_{\rm X, h}$}  &  \scriptsize{$-0.99$ (1.5$\times10^{-4}$)}  &  \scriptsize{$0.98$ (0.02)}  & \scriptsize{$-0.98$ (0.13)}  &  \scriptsize{$0.99$ ($1.3\times10^{-9}$)}   & \scriptsize{$0.30$ (7.2$\times10^{-6}$)}    &  \scriptsize{$-0.386$ (0.39)}  \\
  \normalsize{$F_{\rm R} - F_{\rm 95 GHz}$}  &  \scriptsize{$-$}  & \scriptsize{$-$} & \scriptsize{$0.94$ (0.22)}  &  \scriptsize{$0.99$ ($3.4\times10^{-9}$)}   & \scriptsize{$-$}    &  \scriptsize{$1.00$ (1.9$\times10^{-5}$)}  \\
\normalsize{$F_{\rm R}$ - P}  &  \scriptsize{$-0.91$ (0.01)}   &  \scriptsize{$0.95$ (0.05)} & \scriptsize{$-0.18$ (0.88)}  &  \scriptsize{$-0.97$ ($1.2\times10^{-8}$)}  & \scriptsize{$-0.41$ ($4.1\times 10^{-8}$)}    &  \scriptsize{$0.40$ (0.38)} \\
\normalsize{$F_{\rm R}$ - $\theta$}   & \scriptsize{$-0.82$ (8.9$\times 10^{-4}$)}  & \scriptsize{$-0.97$ (0.03)}  & \scriptsize{$-0.06$ (0.96)}  &  \scriptsize{$0.98$ ($5.9\times10^{-9}$)}  & \scriptsize{$-0.33$ (7.0$\times10^{-7}$)}   &  \scriptsize{$-0.23$ (0.62)}      \\
\normalsize{$\theta$ - P}  & \scriptsize{$0.93$  (5.1$\times10^{-3}$)} & \scriptsize{$-0.92$ (0.08)}  & \scriptsize{$0.99$ (0.05)}  &  \scriptsize{$-0.96$ ($6.7\times 10^{-8}$)}  & \scriptsize{$0.47$ ($5.3\times10^{-8}$)}  &  \scriptsize{$-0.51$ (0.24)}   \\
\normalsize{$F_{\rm R}$ - u}  &  \scriptsize{$-0.37$ (0.46)}  &  \scriptsize{$-0.96$ (0.02)} & \scriptsize{$-0.03$ (0.98)}  &  \scriptsize{$0.97$ ($6.1\times10^{-8}$)}  & \scriptsize{$-0.27$ (1.5$\times10^{-5}$)}   &  \scriptsize{$0.56$ (0.16)}   \\
\normalsize{$F_{\rm R}$ - q}  &  \scriptsize{$-0.84$ (0.04)}  &  \scriptsize{$-0.93$ (0.06)}  & \scriptsize{$-0.73$ (0.46)}  &  \scriptsize{$-0.62$ (1.3$\times10^{-3}$)}  & \scriptsize{$-0.36$ (2.0$\times10^{-8}$)}   &  \scriptsize{$0.11$ (0.83)}  \\
\normalsize{u - q}  &  \scriptsize{$0.11$ (0.85)}  &  \scriptsize{$0.92$ (0.07)} & \scriptsize{$-0.64$ (0.57)}  &  \scriptsize{$-0.73$ (1.6$\times10^{-4}$)}  & \scriptsize{$0.76$ ($3.3\times10^{-9}$)}   &  \scriptsize{$-0.62$ (0.13)} \\
\hline
\hline
\end{tabular}
\end{center}
\begin{center}
$F_{\rm TeV}$, $F_{\rm X, s}$, $F_{\rm X, h}$ and $F_{\rm R}$ are the TeV $\gamma$-ray, soft X-rays, hard-ray and radio fluxes, respectively.  Col (2) May 2008, col (3) March 2010, col (4), col (5), col (6) and col (7) April 2013 for periods 1, 2, 3 and 4.  From col (2) to col (7), numbers in parenthesis are the corresponding p values.\\
\end{center} 
\end{table*}
%%%%%%%%%%%%%%%%%%%%%%%%%%%%%%%%%%%%%%%%%%%%%%%%%%%%%%%%%%%%%%%%%%%%%%%%%
%
\begin{table*}
\begin{center}\renewcommand{\arraystretch}{1.4}\addtolength{\tabcolsep}{2pt}
%\vspace{4cm}
\caption{Intranight variability observations from  2013 April 13 to 19}\label{daily_var}
\begin{tabular}{ c c c c c c c c c c}
\hline
\hline
\normalsize{MJD} & \normalsize{Parameter}& \normalsize{Max}&\normalsize{Min} & \normalsize{Average}& \normalsize{Y(\%)}&\normalsize{$\mu(\%)$}& \normalsize{$\cal F $} & \normalsize{p-value} & \normalsize{$\tau_{\rm \nu, min (R)}$ (hr)} \\
(1)            &   (2)    &      (3)          &      (4)         &      (5)           &   (6)    &    (7)            &  (8)   & (9) & (10)\\    
\hline
\hline
			    & \normalsize{$\Delta m_R$}                          &  \scriptsize{$-$}        &     \scriptsize{$-$}         &       \scriptsize{$-$}      &    \scriptsize{$-$}          &  \scriptsize{$-$}                 &  \scriptsize{$-$}&   \scriptsize{$0.32$}&  \scriptsize{ $-$ }\\
\normalsize{56395} & \normalsize{P(\%)}                         &  \scriptsize{2.30$\pm$0.17}        &     \scriptsize{1.92$\pm$0.12}         &       \scriptsize{$2.32\pm0.05$}      &    \scriptsize{7.71}          &  \scriptsize{2.27}                 &  \scriptsize{0.09} & \scriptsize{$0.79$}&  \scriptsize{$-$}\\
{(April 13)}&\normalsize{$\theta(\degr)$}              &  \scriptsize{104.2$\pm$2.1}        &     \scriptsize{94.2$\pm$2.0}         &       \scriptsize{$97.91\pm0.56$}      &    \scriptsize{11.15}          &  \scriptsize{0.58}                 &  \scriptsize{0.06}& \scriptsize{$0.15$} &  \scriptsize{$-$}\\
\hline
			    & \normalsize{$\Delta m_R$}                          &  \scriptsize{$-$}        &     \scriptsize{$-$}         &       \scriptsize{$-$}      &    \scriptsize{$-$}          &  \scriptsize{$-$}                 &  \scriptsize{$-$} & \scriptsize{$1.1\times 10^{-4}$ (*)}& \scriptsize{  3.30$\pm$0.17}\\
\normalsize{56396} & \normalsize{P(\%)}                         &  \scriptsize{2.19$\pm$0.18}        &     \scriptsize{0.98$\pm$0.50}         &       \scriptsize{$1.31\pm0.06$}      &    \scriptsize{6.52}          &  \scriptsize{4.85}                 &  \scriptsize{0.14} & \scriptsize{$9.2\times 10^{-3}$ (*)}&  \scriptsize{$-$}\\
{(April 14)} &\normalsize{$\theta(\degr)$}              &  \scriptsize{173.4$\pm$4.2}        &     \scriptsize{152.9$\pm$2.3}         &       \scriptsize{$156.94\pm0.99$}      &    \scriptsize{3.48}          &  \scriptsize{0.64}                 &  \scriptsize{0.04}& \scriptsize{$6.1\times 10^{-6}$ (*)} &  \scriptsize{$-$}\\
\hline
			    & \normalsize{$\Delta m_R$}                          &  \scriptsize{$-$}        &     \scriptsize{$-$}         &       \scriptsize{$-$}      &    \scriptsize{$-$}          &  \scriptsize{$-$}                 &  \scriptsize{$-$} & \scriptsize{$1.0\times 10^{-5}$ (*)}& \scriptsize{  2.34$\pm$ 0.12}\\
\normalsize{56397} & \normalsize{P(\%)}                         &  \scriptsize{2.98$\pm$0.16}        &     \scriptsize{1.56$\pm$0.16}         &       \scriptsize{$2.33\pm0.03$}      &    \scriptsize{7.38}          &  \scriptsize{1.29}                 &  \scriptsize{0.39}& \scriptsize{$2.0\times 10^{-4}$ (*)} &  \scriptsize{$-$}\\
{(April 15)}			    &\normalsize{$\theta(\degr)$}              &  \scriptsize{190.1$\pm$1.3}        &     \scriptsize{179.9$\pm$1.2}         &       \scriptsize{$184.63\pm0.36$}      &    \scriptsize{74.57}          &  \scriptsize{0.19}                 &  \scriptsize{0.04}& \scriptsize{$4.0\times 10^{-5}$ (*)} &  \scriptsize{$-$}\\
\hline\hline
			    & \normalsize{$\Delta m_R$}                          &  \scriptsize{$-$}        &     \scriptsize{$-$}         &       \scriptsize{$-$}      &    \scriptsize{$-$}          &  \scriptsize{$-$}                 &  \scriptsize{$-$}&\scriptsize{$5.1\times 10^{-6}$ (*)} & \scriptsize{ 17.82$\pm$0.93} \\
\normalsize{56398} & \normalsize{P(\%)}                         &  \scriptsize{2.73$\pm$0.16}        &     \scriptsize{1.81$\pm$0.16}         &            \scriptsize{$2.55\pm0.03$}      &    \scriptsize{45.58}          &  \scriptsize{1.20}                 &  \scriptsize{0.25}& \scriptsize{$2.1\times 10^{-5}$ (*)} &  \scriptsize{$-$}\\
{(April 16)} &\normalsize{$\theta(\degr)$}              &  \scriptsize{190.1$\pm$2.0}        &     \scriptsize{181.3$\pm$2.3}         &     \scriptsize{$174.69\pm0.31$}      &    \scriptsize{4.81}          &  \scriptsize{0.18}                 &  \scriptsize{0.03}& \scriptsize{$5.1\times 10^{-4}$ (*)} &  \scriptsize{$-$}\\
\hline
			    & \normalsize{$\Delta m_R$}                          &  \scriptsize{$-$}        &     \scriptsize{$-$}         &       \scriptsize{$-$}      &    \scriptsize{$-$}          &  \scriptsize{$-$}                 &  \scriptsize{$-$}&\scriptsize{$0.06$}& \scriptsize{$-$} \\
\normalsize{56399} & \normalsize{P(\%)}                         &  \scriptsize{3.93$\pm$0.16}        &     \scriptsize{2.74$\pm$0.16}         &       \scriptsize{$3.62\pm0.03$}      &    \scriptsize{34.49}          &  \scriptsize{0.94}                 &  \scriptsize{0.18} & \scriptsize{$0.02$} &  \scriptsize{$-$}\\
{(April 17)} &\normalsize{$\theta(\degr)$}              &  \scriptsize{191.0$\pm$1.2}        &     \scriptsize{185.9$\pm$1.4}         &       \scriptsize{$189.23\pm0.26$}      &    \scriptsize{2.44}          &  \scriptsize{0.14}                 &  \scriptsize{0.01} & \scriptsize{$0.23$ } &  \scriptsize{$-$}\\
\hline
			    & \normalsize{$\Delta m_R$}                          &  \scriptsize{$-$}        &     \scriptsize{$-$}         &       \scriptsize{$-$}      &    \scriptsize{$-$}          &  \scriptsize{$-$}                 &  \scriptsize{$-$} &  \scriptsize{$0.68$}& \scriptsize{ $-$ }\\
\normalsize{56400} & \normalsize{P(\%)}                         &  \scriptsize{7.41$\pm$0.17}        &     \scriptsize{5.62$\pm$0.17}         &       \scriptsize{$6.87\pm0.03$}      &    \scriptsize{30.67}          &  \scriptsize{0.43}                 &  \scriptsize{0.15} & \scriptsize{$0.03$} &  \scriptsize{$-$}\\
{(April 18)}  &\normalsize{$\theta(\degr)$}              &  \scriptsize{192.0$\pm$0.9}        &     \scriptsize{183.8$\pm$1.}         &       \scriptsize{$188.34\pm0.15$}      &    \scriptsize{3.47}          &  \scriptsize{0.08}                 &  \scriptsize{0.02} & \scriptsize{$0.66$} &  \scriptsize{$-$}\\
\hline
			    & \normalsize{$\Delta m_R$}                          &  \scriptsize{$-$}        &     \scriptsize{$-$}         &       \scriptsize{$-$}      &    \scriptsize{$-$}          &  \scriptsize{$-$}                 &  \scriptsize{$-$} & \scriptsize{$1.1\times 10^{-7}$ (*)}& \scriptsize{ 5.91$\pm$0.31}\\
\normalsize{56401} & \normalsize{P(\%)}                         &  \scriptsize{4.79$\pm$0.17}        &     \scriptsize{2.61$\pm$0.17}         &       \scriptsize{$3.91\pm0.02$}      &    \scriptsize{61.06}          &  \scriptsize{0.64}                 &  \scriptsize{0.30} & \scriptsize{$5.1\times 10^{-14}$ (*)} &  \scriptsize{$-$} \\
{(April 19)}  &\normalsize{$\theta(\degr)$}              &  \scriptsize{194.7$\pm$1.2}        &     \scriptsize{180.8$\pm$1.4}         &       \scriptsize{$189.27\pm0.18$}      &    \scriptsize{7.36}          &  \scriptsize{0.10}                 &  \scriptsize{0.04} & \scriptsize{$7.0\times 10^{-10}$ (*)}&  \scriptsize{$-$} \\\hline
\hline
\end{tabular}
\end{center}
\begin{center}
Column (2) $\Delta\,m_{R}$ are the R-band differential magnitudes, P(\%) the polarization percentage and $\theta(\deg)$ the position angle. Columns (3), (4) and (5) show the maximum, minimum and average values of the parameters.  Columns (6), (7) and (8) are  the amplitude of the variations, the fluctuation index and the fractional variability index. Column (9) shows the p-values. The (*) show nights where intranight variations are detected at the significance level. The minimum variability timescale obtained for each night is shown in column (10). The cadence ranges from 1 to 5 hours.  \\
\end{center}
\end{table*}
%%%%%%%%%%%%%%%%%%%%%%%%%%%%%%%%%%%%%%%%%%%%%%%%%%%%%%%%%%%%%%%%%%%%%%%%% The parameters shown in columns (6), (7) and (8)  are explained in Table \ref{stat_year}
%
\begin{table*}
\begin{center}\renewcommand{\arraystretch}{1.7}\addtolength{\tabcolsep}{6pt}
%\vspace{4cm}
\caption{Electron spectral indexes and correlation coefficients of the normalized Stokes parameters from  2013 April 13 to 19}\label{nightly_analysis}
\begin{tabular}{ c c c c c c c}
\hline
\hline
\multicolumn{6}{c}{\hspace{6cm}\sf {\large $\alpha_e$ }}\\
\cline{4-6}
\normalsize{MJD} & \normalsize{$m_{q-u}\,\,\,$} ($\chi^2$)& \normalsize{r$_{q-u}$ (p-value)}&\normalsize{$\beta=0.1\,\,\,$($\chi^2$)}&\normalsize{$\beta=0.15\,\,\,$($\chi^2$)}&\normalsize{$\beta=0.20\,\,\,$($\chi^2$)} \\
(1)            &   (2)    &      (3)          &      (4)         &      (5)           &   (6) \\    
\hline
\hline
\normalsize{56395}\,\scriptsize{(April 13)} & \scriptsize{$-0.99\pm0.27\,\,\,$(0.05) }   &  \scriptsize{$-0.65$ (8.1$\times10^{-3}$)}  &    \scriptsize{$5.55\pm0.09\,\,\,$}(7.62)&    \scriptsize{$2.78\pm0.06\,\,\,$}(7.42) &    \scriptsize{$1.46\pm0.04\,\,\,$}(7.47) \\
\normalsize{56396}\,\scriptsize{(April 14)} & \scriptsize{$-0.13\pm0.03\,\,\,$(0.02)}     & \scriptsize{$-0.16$ (0.82)}    &\scriptsize{$3.46\pm0.15\,\,\,$}(4.81) &    \scriptsize{$1.46\pm0.09\,\,\,$}(4.84) &    \scriptsize{$0.54\pm0.06\,\,\,$}(4.84)\\
\normalsize{56397}\,\scriptsize{(April 15)} & \scriptsize{$-0.32\pm0.11\,\,\,$(0.91)}    & \scriptsize{$-0.26$ (0.06)}    &\scriptsize{$5.55\pm0.04\,\,\,$}(93.43) &    \scriptsize{$2.79\pm0.03\,\,\,$}(95.38) &    \scriptsize{$1.47\pm0.02\,\,\,$}(94.14)\\
\normalsize{56398}\,\scriptsize{(April 16)} & \scriptsize{$-0.56\pm0.09\,\,\,$}(045)    & \scriptsize{$-0.72$ (2.5$\times10^{-6}$)} &\scriptsize{$5.96\pm0.05\,\,\,$}(127.0)  &    \scriptsize{$3.05\pm0.03\,\,\,$}(127.8) &    \scriptsize{$1.64\pm0.02\,\,\,$}(129.7)\\
\normalsize{56399}\,\scriptsize{(April 17)} & \scriptsize{$0.59\pm0.22\,\,\,$}(0.14)     &\scriptsize{0.47 (2.8$\times10^{-3}$)} &\scriptsize{$7.63\pm0.05\,\,\,$}(81.6)  &    \scriptsize{$4.13\pm0.03\,\,\,$}(81.95) &    \scriptsize{$2.43\pm0.02\,\,\,$}(78.82)\\
\normalsize{56400}\,\scriptsize{(April 18)} & \scriptsize{$0.27\pm0.08\,\,\,$}(1.02)    & \scriptsize{$0.24$ (0.13)} &\scriptsize{$11.63\pm0.03\,\,\,$}(260.9)  &    \scriptsize{$6.76\pm0.02\,\,\,$}(258.1) &    \scriptsize{$4.35\pm0.02\,\,\,$}(269.8)\\
\normalsize{56401}\,\scriptsize{(April 19)} & \scriptsize{$0.48\pm0.06\,\,\,$}(1.23)    & \scriptsize{$0.74$ ($<10^{-8}$)} &\scriptsize{$8.11\pm0.04\,\,\,$} (546.8) &    \scriptsize{$4.44\pm0.09\,\,\,$}(551.3) &    \scriptsize{$2.66\pm0.02\,\,\,$}(545.5)\\
\hline
\hline
\end{tabular}
\end{center}
\begin{center}
Columns (2) and (3) show the slope and Pearson's correlation coefficients. Columns (4), (5)  and (6) are the average values of the electron power indexes.\\
\end{center}
\end{table*}

%%%%%%%%%%%%%%%%%%%%%%%%%%%%%%%%%%%%%%%%%%%%%%%%%%%%%%%%%%%%%%%%%%%%%%%%%
%
\begin{table*}
\begin{center}\renewcommand{\arraystretch}{1.6}\addtolength{\tabcolsep}{6pt}
%\vspace{4cm}
\caption{Table 8. Variability statistical parameters presented in 2013 April and May }\label{episodes}
\begin{tabular}{c c c c c c c}
\hline
\hline
 \normalsize{Parameter}& \normalsize{Max}&\normalsize{Min} & \normalsize{Average}& \normalsize{Y(\%)}&\normalsize{$\mu(\%)$}& \normalsize{$\cal F $}  \\
 (1)            &   (2)    &      (3)          &      (4)         &      (5)           &   (6)    &    (7) \\    
\hline
\hline
\multicolumn{2}{c}{\sf First period: From 2013 April 10  (MJD 56392) to 13} \\
\cline{1-2}
 \normalsize{$F_{\rm R}$(mJy)}                          &  \scriptsize{93.60$\pm$3.53}        &     \scriptsize{53.82$\pm$2.11}         &       \scriptsize{$69.99\pm0.69$}      &    \scriptsize{$56.72$}          &  \scriptsize{0.99}                 &  \scriptsize{0.27}\\
 \normalsize{P(\%)}                         &  \scriptsize{3.77$\pm$0.17}        &     \scriptsize{1.86$\pm$0.18}         &       \scriptsize{$2.59\pm0.10$}      &    \scriptsize{72.59}          &  \scriptsize{3.87}                 &  \scriptsize{0.34} \\
\normalsize{$\theta(\degr)$}              &  \scriptsize{154.9$\pm$1.1}        &     \scriptsize{137.7$\pm$1.8}         &       \scriptsize{$144.93\pm0.86$}      &    \scriptsize{11.69}          &  \scriptsize{0.59}                 &  \scriptsize{0.06} \\
\hline
\hline
\multicolumn{2}{c}{\sf Second period: From 2013 April 13 (MJD 56395) to 15} \\
\cline{1-2}
 \normalsize{$F_{\rm R}$(mJy)}                          &  \scriptsize{57.34$\pm$2.23}        &     \scriptsize{48.90$\pm$1.98}         &       \scriptsize{$51.64\pm0.19$}      &    \scriptsize{$15.95$}          &  \scriptsize{0.37}                 &  \scriptsize{0.08}\\
 \normalsize{P(\%)}                         &  \scriptsize{2.98$\pm$2.4}        &     \scriptsize{2.19$\pm$0.18 }         &       \scriptsize{$1.97\pm0.04$}      &    \scriptsize{73.42}          &  \scriptsize{2.02}                 &  \scriptsize{0.40} \\
\normalsize{$\theta(\degr)$}              &  \scriptsize{190.1$\pm$2.0}        &     \scriptsize{94.2$\pm$0.16}         &       \scriptsize{$119.48\pm0.50$}      &    \scriptsize{55.40}          &  \scriptsize{0.42}                 &  \scriptsize{0.26} \\
\hline
\hline
\multicolumn{2}{c}{\sf  Third period: From April 15 (MJD 56397) to 19} \\
\cline{1-2}
 \normalsize{$F_{\rm R}$(mJy)}                          &  \scriptsize{55.92$\pm$2.31}        &     \scriptsize{47.00$\pm$2.11}         &       \scriptsize{$51.74\pm0.06$}      &    \scriptsize{$16.89$}          &  \scriptsize{0.12}                 &  \scriptsize{0.09}\\
 \normalsize{P(\%)}                         &  \scriptsize{7.41$\pm$0.17}        &     \scriptsize{1.81$\pm$0.16}         &       \scriptsize{$3.84\pm0.01$}      &    \scriptsize{171.08}          &  \scriptsize{0.33}                 &  \scriptsize{0.68} \\
\normalsize{$\theta(\degr)$}              &  \scriptsize{194.7$\pm$1.7}        &     \scriptsize{185.9$\pm$1.4}         &       \scriptsize{$185.63\pm0.11$}      &    \scriptsize{13.48}          &  \scriptsize{0.06}                 &  \scriptsize{0.07} \\
\hline
\hline
\multicolumn{2}{c}{\sf  From 2013 May 10 (MJD 56422) to 17 (MJD 56429) } \\
\cline{1-2}
 \normalsize{$F_{\rm R}$(mJy)}                          &  \scriptsize{64.49$\pm$2.49}        &     \scriptsize{47.77$\pm$1.91}         &       \scriptsize{$53.02\pm0.35$}      &    \scriptsize{$31.33$}          &  \scriptsize{0.65}                 &  \scriptsize{0.15}\\
 \normalsize{P(\%)}                         &  \scriptsize{11.00$\pm$0.18}        &     \scriptsize{7.16$\pm$0.19}         &       \scriptsize{$8.92\pm0.07$}      &    \scriptsize{42.85}          &  \scriptsize{0.80}                 &  \scriptsize{0.21} \\
\normalsize{$\theta(\degr)$}              &  \scriptsize{217.8$\pm$0.8}        &     \scriptsize{190.9$\pm$0.8}         &       \scriptsize{$200.59\pm0.33$}      &    \scriptsize{13.39}          &  \scriptsize{0.16}                 &  \scriptsize{0.07} \\
\hline
\end{tabular}
\end{center}
\begin{center}
The parameters Y(\%), $\mu(\%)$  and $\cal F $ are  the amplitude of the variations, the fluctuation index and the fractional variability index. \\
\end{center}
\end{table*}
%
%%%%%%%%%%%%%%%%%%%%%%%%%%%%%%%%%%%%%%%%%%%%%%%%%%%%%%%%%%%%%%%%%%%%%%%%%%
%
\begin{table*}
\begin{center}\renewcommand{\arraystretch}{1.3}\addtolength{\tabcolsep}{1pt}
\caption{leptonic model parameters}\label{sed_parameters}
\begin{tabular}{ l c c c c c c c c c}
\hline
\hline
\normalsize{} & \scriptsize{ MJD 54590}&  \scriptsize{ MJD 54591}&  \scriptsize{ MJD 54594} &  \scriptsize{ MJD 55268} &  \scriptsize{ MJD 55270} &  \scriptsize{ Quiescent  1} &  \scriptsize{ Quiescent 2} \\
\normalsize{} & \scriptsize{(2008 May 04)}&  \scriptsize{(2008 May 05)}&  \scriptsize{(2008 May 08)} &  \scriptsize{(2010 March 12)} &  \scriptsize{(2010 March 14)} &  \scriptsize{ } &  \scriptsize{ } \\

\hline
\hline
\multicolumn{2}{c}{Obtained quantities} \\
\cline{1-2}
\scriptsize{$\delta_D$} & \scriptsize{24} & \scriptsize{24}&  \scriptsize{24}&  \scriptsize{22}&  \scriptsize{22}&  \scriptsize{20}&  \scriptsize{20} \\
\scriptsize{$B$ ($\times 10^{-3}$ G)} & \scriptsize{55.3}&  \scriptsize{54.6}&  \scriptsize{54.9}&  \scriptsize{43.3}&  \scriptsize{43.1}&  \scriptsize{34.3}&  \scriptsize{34.1} \\
\scriptsize{$r_d$ ($\times 10^{16}$ cm)} & \scriptsize{2.5} & \scriptsize{2.5}   & \scriptsize{2.5}& \scriptsize{4.3}& \scriptsize{4.3}& \scriptsize{5.0}& \scriptsize{5.0}  \\
\scriptsize{$N_e$ ($\times 10^{-1}$ cm$^{-3}$)} & \scriptsize{1.9} &   \scriptsize{1.7}&  \scriptsize{1.6} &  \scriptsize{2.6} &  \scriptsize{2.3} &  \scriptsize{1.2} &  \scriptsize{1.0} \\\hline
\multicolumn{2}{c}{Derived quantities} \\
\cline{1-2}
\scriptsize{$\gamma_{c}$ ($\times 10^5$) } & \scriptsize{2.7}&  \scriptsize{2.8}&  \scriptsize{2.8}&  \scriptsize{2.6}&  \scriptsize{2.6}&  \scriptsize{3.6}&  \scriptsize{3.7} \\
\scriptsize{$\gamma_{c1}$ ($\times 10^4$) } & \scriptsize{5.8}&  \scriptsize{5.8}&  \scriptsize{5.8}&  \scriptsize{2.7}&  \scriptsize{2.7}&  \scriptsize{7.8}&  \scriptsize{7.8} \\
\scriptsize{$\gamma_{c2}$ ($\times 10^5$) } & \scriptsize{3.1}&  \scriptsize{3.1}&  \scriptsize{3.1}&  \scriptsize{2.8}&  \scriptsize{2.8}&  \scriptsize{4.0}&  \scriptsize{4.0} \\
\scriptsize{$\gamma_{e,max}$ ($\times 10^8$) } & \scriptsize{1.2}&  \scriptsize{1.2}&  \scriptsize{1.2}&  \scriptsize{1.4}&  \scriptsize{1.4}&  \scriptsize{1.6}&  \scriptsize{1.6} \\
\scriptsize{$U_e/U_B$\,\, ($\times 10^1$)}   & \scriptsize{1.1}&  \scriptsize{1.0}&  \scriptsize{0.9}&  \scriptsize{2.2}&  \scriptsize{2.0}&  \scriptsize{1.7}&  \scriptsize{1.4} \\
\scriptsize{$L_e$  ($\times 10^{43}$ erg/s)} & \scriptsize{1.7}&  \scriptsize{1.5}&  \scriptsize{1.4}&  \scriptsize{5.7}&  \scriptsize{5.1}&  \scriptsize{3.0}&  \scriptsize{2.5} \\
\scriptsize{$L_B$  ($\times 10^{42}$ erg/s)} & \scriptsize{4.1}&  \scriptsize{4.0}&  \scriptsize{4.1}&  \scriptsize{6.3}&  \scriptsize{6.2}&  \scriptsize{4.4}&  \scriptsize{4.4} \\
\scriptsize{$L_p$  ($\times 10^{43}$ erg/s)} & \scriptsize{1.0}&  \scriptsize{0.9}&  \scriptsize{0.8}&  \scriptsize{3.3}&  \scriptsize{2.9}&  \scriptsize{1.7}&  \scriptsize{1.4} \\
\scriptsize{$L_{\rm jet}$  ($\times 10^{43}$ erg/s)} & \scriptsize{3.1}&  \scriptsize{2.8}&  \scriptsize{2.6}&  \scriptsize{9.6}&  \scriptsize{8.6}&  \scriptsize{5.1}&  \scriptsize{4.4} \\
\hline
\hline
\end{tabular}
\end{center}

{\scriptsize 
$\delta_D$ is the Doppler factor,  $B$ is the magnetic field,  $r_d$ is the size of the emitting radius,  $N_e$ is the electron number density,   $\gamma_{c}$  is the electron cooling Lorentz factor,   $\gamma_{c1}$,  $\gamma_{c2}$ and $\gamma_{e,max}$ are the electron Lorentz factors for breaks (1 and 2) and maximum. $U_e$ and $U_B$ are the densities carried by electrons and magnetic field. $L_e$,  $L_B$ and $L_p$ are the electron, magnetic field and proton luminosities, and $L_{\rm jet}$ is the total jet power. Quiescent state 1:  from 2008 August 05 (MJD 54683) to  2009 June 18 (MJD 55000).  Quiescent state 2:  from 2010 November 16 (MJD 55516) to 2012 June 28 (MJD 56106).}

\end{table*}
%
%%%%%%%%%%%%%%%%%%%%%%%%%%%%%%%%%%%%%%%%%%%%%%%%%%%%%%%%%%%%%%%%%%%%%%%%%%
%%%%%%%%%%%%%%%%%%%%%%%%%%%%%%%%%%%%%%%%%%%%%%%%%%%%%%%%%%%%%%%%%%%%%%%%%%
%%%%%%%%%%%%%%%%%%%%%%%%%%%%%%%%%%%%%%%%%%%%%%%%%%%%%%%%%%%%%%%%%%%%%%%%%%
%%%%%%%%%%%%%%%%%%%%%%%%%%%%%%%%%%%%%%%%%%%%%%%%%%%%%%%%%%%%%%%%%%%%%%%%%%
%
%\clearpage
%
%
%%%%%%%%%%%%%%%%%%%%%%%%%%%%%%%%%%%%%%%%%%%%%%%%%%%%%%%%%%%%%%%%%%%%%%%%%%
%
%\clearpage
%%%%%%%%%%%%%%%%%%%%%%%%%%%%%%%%%%%%%%%%%%%%%%%%%%%%%%%%%%%%%%%%%%%%%%%%%%
%
\clearpage
%
%%%%%%%%%%%%%%%%%%%%%%%%%%%%%%%%%%%%%%%%%%%%%%%%%%%%%%%%%%%%%%%%%%%%%%%%%%%
%
%%%%%%%%%%%%%%%%%%%%%%%%%%%%%%%%%%%%%%%%%%%%%%%%%%%%%%%%%%%%%%%%%%%%%%%%%%
%
%\clearpage
%%%%%%%%%%%%%%%%%%%%%%%%%%%%%%%%%%%%%%%%%%%%%%%%%%%%%%%%%%%%%%%%%%%%%%%%%%
%
%\clearpage
%%%%%%%%%%%%%%%%%%%%%%%%%%%%%%%%%%%%%%%%%%%%%%%%%%%%%%%%%%%%%%%%%%%%%%%%%%
%
%
%\clearpage

%\clearpage
%
%%%%%%%%%%%%%%%%%%%%%%%%%%%%%%%%%%%%%%%%%%%%%%%%%%%%%%%%%%%%%%%%%%%%%%%%%%
%
%\clearpage
%%%%%%%%%%%%%%%%%%%%%%%%%%%%%%%%%%%%%%%%%%%%%%%%%%%%%%%%%%%%%%%%%%%%%%%%%%%
%
%
%\clearpage
%

%%%%%%%%%%%%%%%%%%%%%%%%%%%%%%%%%%%%%%%%%%%%%%%%%%%%%%%%%%%%%%%%%%%%%%%%%%
%
%\clearpage
%
%%%%%%%%%%%%%%%%%%%%%%%%%%%%%%%%%%%%%%%%%%%%%%%%%%%%%%%%%%%%%%%%%%%%%%%%%%
%
%%%%%%%%%%%%%%%%%%%%%%%%%%%%%%%%%%%%%%%%%%%%%%%%%%%%%%%%%%%%%%%%%%%%%%%%%%
%
%%%%%%%%%%%%%%%%%%%%%%%%%%%%%%%%%%%%%%%%%%%%%%%%%%%%%%%%%%%%%%%%%%%%%%%%%%

\clearpage
%%%%%%%%%%%%%%%%%%%%%%%%%%%%%%%%%%%%%%%%%%%%%%%%%%%%%%%%%%%%%%%%%%%%%%%%%%
%
%
%%%%%%%%%%%%%%%%%%%%%%%%%%%%%%%%%%%%%%%%%%%%%%%%%%%%%%%%%%%%%%%%%%%%%%%%%%%
%
%\clearpage
%
%OJO NISSIM, EN LA ULTIMA FIGURA NO VEO OS PUNTOS NEGROS QUE REPRESENTAN A  LA BANDA R COMO DICE EL PIE DE FIGURA.....

\end{document}